\documentstyle[12pt]{article}

\def\ar{\rightarrow}
\def\bib{\bibitem}
\def\dim{\,\mbox{dim}\,}
\def\Det{\,\mbox{Det}\,}

\def\Dsl{D\!\!\!\!/}
\def\intx{\int\! d^{\sl 4}x}
\def\intX{\int\! d^{\sl 4}X\,}
\def\intp{\int\! \frac{d^{\sl 4}p}{(2{\pi})^4}}
\def\intP{\int\! \frac{d^{\sl 4}P}{(2{\pi})^4}\,}

\def\intr{\int\!\!\!\!\!\!_{^{reg}}}
\def\lar{\longrightarrow}
\def\Ln{\mbox{Ln}}

\def\pa{\partial}
\def\rvec{\!\!\!\!^{^\rightarrow}}
\def\lvec{\!\!\!\!^{^\leftarrow}}
\def\TA{\:\bar{\!\!A}}

\def\Tr{\,\mbox{Tr}\,}

\def\al{\alpha}
\def\be{\beta}
\def\ga{\gamma}
\def\de{\delta}
\def\ep{\varepsilon}
\def\ze{\zeta}

\def\la{\lambda}
\def\va{\varphi}
\def\si{\sigma}
\def\om{\omega}

\def\Ga{{\it\Gamma}}
\def\La{{\it\Lambda}}
\def\Om{{\it\Omega}}

\def\Pit{\!{\it\Pi}}
\def\Pitc{\!{\hat{\it\Pi}^*}}
\def\Pitt{{\hat{\it\Pi}}}
\def\Pitch{\!{\tilde{\hat{\it\Pi}^*}}}
\def\Pitth{{\tilde{\hat{\it\Pi}}}}

\def\TPit{\bar{\Pit}}
\def\Si{\Sigma}

\def\beq{\begin{equation}}
\def\eeq{\end{equation}}
\def\bed{\begin{displaymath}}
\def\eed{\end{displaymath}}
\def\beqq{\begin{eqnarray}}
\def\eeqq{\end{eqnarray}}
\def\bedd{\begin{eqnarray*}}
\def\eedd{\end{eqnarray*}}

\begin{document}

\centerline{\normalsize\bf III - CONSERVATION OF GRAVITATIONAL ENERGY MOMENTUM} \centerline{\normalsize\bf AND RENORMALIZABLE QUANTUM THEORY OF GRAVITATION}

\vspace*{0.9cm}
\centerline{\footnotesize C. WIESENDANGER}
\baselineskip=12pt
\centerline{\footnotesize\it Aurorastr. 24, CH-8032 Zurich}
\centerline{\footnotesize E-mail: christian.wiesendanger@ubs.com}

\vspace*{0.9cm}
\baselineskip=13pt
\abstract{Viewing gravitational energy-momentum $p_G^\mu$ as equal by observation, but different in essence from inertial energy-momentum $p_I^\mu$ naturally leads to the gauge theory of volume-preserving diffeormorphisms of an inner Minkowski space ${\bf M\/}^{\sl 4}$ which can describe gravitation at the classical level. This theory is quantized in the path integral formalism starting with a non-covariant Hamiltonian formulation with unconstrained canonical field variables and a manifestly positive Hamiltonian. The relevant path integral measure and weight are then brought into a Lorentz- and gauge-covariant form allowing to express correlation functions - applying the De Witt-Faddeev-Popov approach - in any meaningful gauge. Next the Feynman rules are developed and the quantum effective action at one loop in a background field approach is renormalized which results in an asymptotically free theory without presence of other fields and in a theory without asymptotic freedom including the Standard Model (SM) fields. Finally the BRST apparatus is developed as preparation for the renormalizability proof to all orders and a sketch of this proof is given.}

\normalsize\baselineskip=15pt

\section{Introduction}
In \cite{chw1} we have started to explore the consequences of viewing the gravitational energy-momentum $p_G^\mu$ as different by its very nature from the inertial energy-momentum $p_I^\mu$, accepting their observed numerical equality as accidential.

As both are conserved this view has led us to look for two different symmetries which through Noether's theorem generate two different conserved four vectors - one symmetry obviously being space-time translation invariance yielding the conserved inertial energy-momentum $p_I^\mu$ vector. To generate an additional conserved four-vector the field concept has proven to be crucial as only fields can carry the necessary inner degrees of freedom to allow for representations of additional inner symmetry groups - in our case an inner translation group yielding the conserved gravitational energy-momentum vector $p_G^\mu$.

Gauging this inner translation group has then naturally led to the gauge field theory of volume-preserving diffeomorphisms of ${\bf M\/}^{\sl 4}$, at the classical level, thereby generalizing the Yang-Mills approach for compact Lie groups acting on a finite number of inner field degrees of freedom (also see \cite{chwA,chwB} for the mathematical framework). The resulting theory is a consistent classical gauge theory and its gauge fields can be coupled in a universal way to any other field.

In \cite{chw2} we then have interpreted the theory as a theory of gravitation reducing the full gauge theory of volume-preserving diffeomorphisms of ${\bf M}^{\sl 4}$ to a gauge theory of its Poincar\'e subgroup $POIN\,{\bf M}^{\sl 4}$. As a consequence of this reduction we have obtained a relativistic description of gravitational fields interacting with point-particle matter and of matter moving in gravitational fields which - after numerical identification of gravitational and inertial energy-momentum and angular-momentum - in the non-relativistic limit has yielded Newton's inverse square law for gravity.

For the theory's viability there remains the problem of a consistent quantization. Not only will we have to deal with the usual short distance divergencies of space-time integrals in a perturbation expansion \cite{stw1,stw2}, but due to the non-compactness of the gauge group we will face additional divergent integrals over inner space which have to be regularized in a way respecting the relevant symmetries (inner Lorentz and scale invariance) - generalizing thereby the finite sums over structure constants appearing in the perturbation series for the Yang-Mills case to the present one.

The solution is related to noting that the classical gauge theory of volume-preserving diffeomorphisms of ${\bf M}^{\sl 4}$ has a linearly realized inner scale invariance which necessarily has to be a symmetry of the quantum effective action as well \cite{stw2}. This suggests a regularization scheme of the divergent integrals over the gauge group respecting inner scale invariance which will yield a renormalizable quantum field theory uniquely determined up to inner rescalings.

Technically we will quantize in the path integral formalism starting with a Hamiltonian formulation of the theory with unconstrained, though neither Lorentz- nor gauge-covariantly looking canonical field variables and a manifestly positive Hamiltonian. Over various steps the relevant path integral measure and weight are brought into a Lorentz- and gauge-covariant form allowing us to express correlation functions first in the Minkowski-plus-axial gauge and - applying the De Witt-Faddeev-Popov approach - in any meaningful gauge. Next the Feynman rules are developed and the quantum effective action at one loop in a background field approach is renormalized which results in an asymptotically free theory without presence of other fields and in a theory without asymptotic freedom if including the Standard Model (SM) fields. Finally the BRST apparatus is developed as preparation for the renormalizability proof to all orders and a sketch of this proof is given.

The notations and conventions used follow closely those of Steven Weinberg in his classic account on the quantum theory of fields \cite{stw1,stw2}. They are presented in the Appendix.

\section{Quantization in the Minkowski-plus-axial Gauge}
In this section we quantize the gauge theory of volume-preserving diffeomorphisms of ${\bf M}^{\sl 4}$ starting with a Hamiltonian formulation of the classical theory equivalent to its Lagrangian formulation in the Minkowski-plus-axial gauge. This allows us to express all quantum amplitudes of interest as path integrals over unconstrained canonical field variables which live in certain functional spaces ensuring the positivity of the Hamiltonian. These integrals look neither Lorentz- nor gauge-invariant. We then show that they can be transformed into explicitly Lorentz- and gauge-invariant expressions to be evaluated in the Minkowski-plus-axial gauge resulting in a ghost-free, covariant and unitary quantum field theory with a positive field energy operator.

Our starting point is the Hamiltonian formulation of the classical gauge theory of volume-preserving diffeomorphisms of ${\bf M}^{\sl 4}$ in terms of a minimal set of unconstrained canonical field variables and with a manifestly positive Hamiltonian as developed in \cite{chw1}. This formulation specifies the physical field content of the theory and comes - after quantization - along with a positive-definite metric in the Hilbert space of state vectors and a manifestly positive energy operator, hence yielding a viable quantum theory. On the other hand it obscures the Lorentz- and gauge invariance embedded in the Lagrangian formulation and comes at the price of non-local relations between the unconstrained field variables and the covariant ones of the Lagrangian formulation.

The independent canonical field variables of the theory are
${\tilde{\hat A}}_{i\,a} (x,K)$ together with their conjugate field variables 
$\Pitch_{j\,b} (x,K)$, defined on the product of space-time and an inner momentum Minkowski space ${\bf M}^{\sl 4}\times {\bf M}^{\sl 4}$, where $i,j = {\sl 1,2}$ and $a,b = {\sl 1,2,3}$ - in total twelve field variables without constraints apart from the reality conditions ${\tilde{\hat A}}_{i\,a} (x,-K)={\tilde{\hat A^*}}_{i\,a} (x,K)$, $\Pitth_{j\,b} (x,-K)=\Pitch_{j\,b} (x,K)$.

The Hamiltonian $H = \int\! d^{\sl 3}x \int\! d^{\sl 4}K\, \La^4\, {\cal H}$ specifying the field dynamics is given in terms of the Hamiltonian density \cite{chw1}
\beqq \label{1} 
{\cal H} &=& \frac{1}{2\,\La^2}\, \sum_{a={\sl 1}}^{\sl 2} \pa_{\sl 3}{\tilde{\hat A^*}}_{{\sl 0}\,a} \cdot \pa_{\sl 3}{\tilde{\hat A}}_{{\sl 0}\,a} 
+ \frac{1}{2}\, \sum_{i;a={\sl 1}}^{\sl 2} \Pitch_{i\,a} \cdot \Pitth_{i\,a}  
\nonumber \\
& &+ \frac{1}{4\,\La^2}\, \sum_{i,j;a={\sl 1}}^{\sl 2} {\tilde{\hat F^*}}_{ij\,a} \cdot {\tilde{\hat F}}_{ij\,a}
+ \frac{1}{2\,\La^2}\, \sum_{i;a={\sl 1}}^{\sl 2}
\pa_{\sl 3}{\tilde{\hat A^*}}_{i\,a}\cdot \pa_{\sl 3}{\tilde{\hat A}}_{i\,a} \nonumber \\
&+& \left(\frac{- K^2}{(K_{\sl 0})^2}\right) \Bigg\{
\frac{1}{2\,\La^2}\, \pa_{\sl 3}{\tilde{\hat A^*}}_{{\sl 0\,3}} \cdot \pa_{\sl 3}{\tilde{\hat A}}_{{\sl 0\,3}} 
+ \frac{1}{2}\, \sum_{i={\sl 1}}^{\sl 2} \Pitch_{i\,{\sl 3}} \cdot \Pitth_{i\,{\sl 3}} \\
& &+ \frac{1}{4\,\La^2}\, \sum_{i,j={\sl 1}}^{\sl 2} {\tilde{\hat F^*}}_{ij\,{\sl 3}} \cdot {\tilde{\hat F}}_{ij\,{\sl 3}}
+ \frac{1}{2\,\La^2}\, \sum_{i={\sl 1}}^{\sl 2}
\pa_{\sl 3}{\tilde{\hat A^*}}_{i\,{\sl 3}}\cdot \pa_{\sl 3}{\tilde{\hat A}}_{i\,{\sl 3}} \Bigg\} \nonumber \\
&\geq& 0. \nonumber
\eeqq

To ensure its positivity the field variables have support on ${\bf M}^{\sl 4}\times \Big({\bf V^+}(K)\cup {\bf V^-}(K)\Big)$, where
\beq \label{2}
{\bf V^\pm}(K) = \{K\in {\bf M^{\sl 4}}\mid -K^2 \geq 0,\: \pm K^{\sl 0} \geq 0\}
\eeq
denote the forward and backward light cones in inner momentum space. The non-local functionals ${\tilde{\hat A}}_{{\sl 0}\,a} \left[{\tilde{\hat A}}_{k\,c},\, \Pitch_{h\,d}\right], {\tilde{\hat F}}_{ij\,a} \left[{\tilde{\hat A}}_{k\,c}\right]$ of the independent canonical variables with $i,j,k,h = {\sl 1,2}$ and $a,c,d = {\sl 1,2,3}$ also have support on ${\bf M}^{\sl 4}\times \Big({\bf V^+}(K)\cup {\bf V^-}(K) \Big)$. Their explicit functional form is specified in Eqns.(\ref{18}) and (\ref{19}) below. $\La$ is a parameter which carries dimension of length ensuring the dimensionlessness of all expressions when counting dimensions w.r.t. inner space.

The energy of a field is positive if its support is limited to the forward and backward light cones Eqn.(\ref{2}) in inner space - a condition which after expanding the fields into free waves in inner space is equivalent to the more physical one that for all free wave states the square of the gravitational energy-momentum vector $K$ which equals the invariant mass squared $-K^2 = M^2 \geq 0$ is positive.

Finally we note that singling out the $a = {\sl 3}$-field components is a pure matter of convention - we could have as well singled out the $a = {\sl 1}$- or $a = {\sl 2}$-field components as will become clear further down.

So $H$ is manifestly positive definite and specifies a consistent classical field dynamics in terms of the regular equal-time Poisson brackets 
\beq \label{3} 
\left\{
{\tilde{\hat A}}_{i\,a} (x,K),\, \Pitch\,\!_{j\,b} (y,Q) \right\}_{x^{\sl 0} = y^{\sl 0}}
= \de_{ij}\, \de_{ab}\, \La^{-4}\, \de^{\sl 4}(K - Q)\, \de^{\sl 3}({\underline x} - {\underline y})
\eeq
for the unconstrained canonical field variables ${\tilde{\hat A}}_{i\,a}$ and $\Pitch_{j\,b}$, as discussed in \cite{chw1}. Note that the classical dynamics corresponding to the Poisson brackets above is consistent with the support condition on the fields and that quantization of these Poisson brackets gives us a Hilbert space with positive definite metric for the quantum states.

${\cal H}$ together with a Hamiltonian density ${\cal H}_M = \sum_n \pi_n\cdot \pa_{\sl 0} \psi_n - {\cal L}_M$ for generic "matter" fields $\psi_m (x,X)$ with conjugates $\pi_n (x,X)$ is our starting point for the path integral quantization. Note that the unconstrained gauge field variables we start with are defined on $K$-space to manifestly implement the support condition on the fields ensuring a positive Hamiltonian and not on $X$-space on which the "matter" fields are defined from the outset for convenience.

The Green functions of the quantized theory are defined as unconstrained path integrals over ${\tilde{\hat A}}_{i\,a} (x,K)$, $\Pitch_{j\,b} (x,K)$, $\psi_m (x,X)$, $\pi_n(x,X)$ with gauge and matter field measures 
\beq \label{4}
\Pi_{\!\!\!\!\!\!_{_{_{x,X;m}}}}\!\!\!\!d\psi_m \cdot
\Pi_{\!\!\!\!\!\!_{_{_{x,K;i={\sl 1},{\sl 2},a={\sl 1},{\sl 2},{\sl 3}}}}} \!\!\!\!\!\!\!\!\!\!\!\!\!d{\tilde{\hat A}}_{i\,a}
\cdot \Pi_{\!\!\!\!\!\!_{_{_{x,X;n}}}}\!\!\!\!d\pi_n \cdot \Pi_{\!\!\!\!\!\!_{_{_{x,K;j={\sl 1},{\sl 2},b={\sl 1},{\sl 2},{\sl 3}}}}}
\!\!\!\!\!\!\!\!\!\!\!\!\!\!\!d\Pitch_{j\,b}
\eeq
and weight
\beqq \label{5}
& & \exp\,i\,\int\left\{\frac{1}{\La}\, \sum_{i,a ={\sl 1}}^{\sl 2} \Pitch_{i\,a}\cdot \pa_{\sl 0} {\tilde{\hat A}}_{i\,a}
- \frac{K^2}{(K_{\sl 0})^2} \frac{1}{\La}\, \sum_{i ={\sl 1}}^{\sl 2} 
\Pitch_{i\,{\sl 3}}\cdot \pa_{\sl 0} {\tilde{\hat A}}_{i\,{\sl 3}}
- {\cal H} \right\} \nonumber \\
& & \quad\quad\quad\quad\quad\quad\quad
\cdot \exp\,i\,\int\left\{\sum_n \pi_n\cdot \pa_{\sl 0} \psi_n - {\cal H}_M \right\}.
\eeqq

Through a series of canonical field transformations and Gaussian integrations we next turn these unconstrained, but neither Lorentz- nor gauge-covariantly looking path integrals into Lorentz- and gauge-covariant ones.

The first step is to bring ${\cal H}$ into a form symmetric in all three $a$-indices, which, however, will obscure the positivity of the Hamiltonian. It is related to the $\sl{3}\times \sl{3}$-matrix
\beq \label{6}
M_{ab} (K) \equiv \de_{ab}  - \frac{K_a\, K_b}{(K_{\sl 0})^2}
\eeq
which is real and symmetric with eigenvalues $1$, $1$ and $-\frac{K^2}{(K_{\sl 0})^2}$. Here $K^2 = -(K_{\sl 0})^2 + \sum_{a={\sl 1}}^{\sl 3}(K_a)^2$ is the Minkowski square.

Because $M (K)$ is symmetric there exists an orthogonal $3\times 3$-matrix $D (K)$, $D^T = D^{-1}$ such that $D^T\, M\, D = \mbox{diag} (1,1,-\frac{K^2}{(K_{\sl 0})^2})$. Rotating the field variables
\beq \label{7}
{\hat A}_i\,^a (K) \equiv
D^a\,_b (K)\, {\tilde{\hat A}}_i\,^b (K) 
\eeq
and using the same transformation for all the terms appearing in Eqn.(\ref{1}) we can rewrite the Hamiltonian density in the symmetric form
\beqq \label{8} 
\!\!\!\!\!\!\!\!\!{\cal H} &=& \frac{1}{2\,\La^2}\, \pa_{\sl 3}{\hat A^*}_{{\sl 0}\,a} \cdot M^{ab}(K)\, \pa_{\sl 3}{\hat A}_{{\sl 0}\,b}
+ \frac{1}{2}\, \sum_{i={\sl 1}}^{\sl 2} \Pitc_{i\,a} \cdot M^{ab}(K)\, \Pitt_{i\,b} \nonumber \\
\!\!\!\!\!\!\!\!\!&+& \frac{1}{4\,\La^2}\, \sum_{i,j={\sl 1}}^{\sl 2} {\hat F^*}_{ij\,a} \cdot M^{ab}(K)\, {\hat F}_{ij\,b} + \frac{1}{2\,\La^2}\, \sum_{i={\sl 1}}^{\sl 2}
\pa_{\sl 3}{\hat A^*}_{i\,a}\cdot M^{ab}(K)\, \pa_{\sl 3}{\hat A}_{i\,b} \\
\!\!\!\!\!\!\!\!\!&\geq& 0 \nonumber,
\eeqq
where we have applied the usual sum convention for $a,b$. Note that starting with the $a={\sl 1}$- or $a={\sl 2}$-field component singled out instead of the $a={\sl 3}$-field components in the definition of the Hamiltonian Eqn.(\ref{1}) above and $M$ accordingly diagonalized as $\mbox{diag} (-\frac{K^2}{(K_{\sl 0})^2},1,1)$ or $\mbox{diag} (1,-\frac{K^2}{(K_{\sl 0})^2},1)$ respectively it becomes obvious that the support for {\it all} field components has to be restricted to ${\bf V^+}(K)\cup {\bf V^-}(K)$ to ensure positivity of the Hamiltonian.

Due to the orthogonality of $D (K)$ resulting in $\det D (K) = 1$ the gauge and "matter" field measures transform into
\beq \label{9}
\Pi_{\!\!\!\!\!\!_{_{_{x,X;m}}}}\!\!\!\!d\psi_m \cdot
\Pi_{\!\!\!\!\!\!_{_{_{x,K;i={\sl 1},{\sl 2},a={\sl 1},{\sl 2},{\sl 3}}}}} \!\!\!\!\!\!\!\!\!\!\!\!\!d{\hat A}_{i\,a}
\cdot \Pi_{\!\!\!\!\!\!_{_{_{x,X;n}}}}\!\!\!\!d\pi_n \cdot \Pi_{\!\!\!\!\!\!_{_{_{x,K;j={\sl 1},{\sl 2},b={\sl 1},{\sl 2},{\sl 3}}}}}
\!\!\!\!\!\!\!\!\!\!\!\!\!\!\!d\Pitc_{j\,b}
\eeq
and the weight into
\beqq \label{10}
& & \exp\,i\,\int\left\{\frac{1}{\La}\, \sum_{i ={\sl 1}}^{\sl 2} \Pitc_{i\,a}\cdot M^{ab}(K)\, \pa_{\sl 0} {\hat A}_{i\,b}
- {\cal H} \right\} \nonumber \\
& & \quad\quad \cdot \exp\,i\,\int\left\{\sum_n \pi_n\cdot \pa_{\sl 0} \psi_n - {\cal H}_M \right\}.
\eeqq

The second step is to Fourier-transform all the fields in inner space which is a canonical transformation. Omitting the $x$-coordinates for notational simplicity we have
\beq \label{11}
A_i\,^a (X) = \int\! \frac{d^{\sl 4}K}{(2{\pi})^2}\, \La^4\,
e^{i\, K\cdot X}\, {\hat A}_i\,^a (K), 
\eeq
where the reality condition on ${\hat A}_i\,^a$
\beq \label{12}
{\hat A}_i\,^a (-K) = {\hat A^*}_i\,^a (K)
\Rightarrow A_i\,^a (X) = A^*_i\,^a (X) 
\eeq 
ensures that the fields over $X$-space are real.

Note that the support condition on ${\hat A}_i\,^a (K)$ does not translate into a simple condition on $A_i\,^a (X)$ which is the reason why we had to start with Fourier-transformed fields to uncover a condition sufficient for the Hamiltonian to be positive. The appropriate functional spaces over which path integrals are to be evaluated are then defined such that their Fourier-transformed elements have support on ${\bf V^+}(K)\cup {\bf V^-}(K)$.

Fourier-transformation of a typical term in Eqn.(\ref{8}) yields 
\beqq \label{13}
\!\!\!\!\!\!\!\!\!\!\!\!& & \int\! d^{\sl 4}K\, \La^4\, \pa_{\sl 3}{\hat A^*}_i\,^a\cdot M_{ab} (K) \, \pa_{\sl 3}{\hat A}_i\,^b = \intX \La^{-4}\, \Bigg\{\pa_{\sl 3}A_i\,^a\cdot \pa_{\sl 3}A_{i\,a} \nonumber \\
\!\!\!\!\!\!\!\!\!\!\!\!& &\quad -\, \pa_{\sl 3}\, \frac{1}{\nabla_{\sl 0}}\, \nabla_a A_i\,^a\cdot \pa_{\sl 3}\, \frac{1}{\nabla_{\sl 0}}\, \nabla_b A_i\,^b \Bigg\}
=\intX \La^{-4}\, \pa_{\sl 3}A_i\,^\al \cdot \pa_{\sl 3}A_{i\,\al},
\eeqq
where we have introduced
\beqq  \label{14}
A_i\,^{\sl 0} (x,X) &\equiv& -\int^{X^{\sl 0}}\!\!\! dS\, \nabla_a A_i\,^a (x;S\, ,X^{\sl 1},X^{\sl 2},X^{\sl 3}) \nonumber \\
&=& -\,\frac{1}{\nabla_{\sl 0}}\, \nabla_a A_i\,^a,
\:\: i = {\sl 1,2}
\eeqq
which is an additional field defined as a functional of the canonical variables $A_i\,^a (x;X)$.

We formally take $A_i\,^{\sl 0}$ as the zero component of a four vector in inner space. As a consequence the $A_i\,^\al$ fulfil the unimodularity constraints in inner space
\beq  \label{15}
\nabla_\al A_i\,^\al = \nabla_{\sl 0} A_i\,^{\sl 0} + \nabla_a A_i\,^a
= 0,\:\: i = {\sl 1,2}, 
\eeq
and analogously for $\Pit_j\,\!^\be$
\beq  \label{16}
\Pit_j\,\!^{\sl 0} (x,X) = -\frac{1}{\nabla_{\sl 0}}\,
\nabla_a \Pit_j\,\!^a ,\:\: i = {\sl 1,2}.
\eeq

As a result we can re-write the Hamiltonian density in a form which is Lorentz-invariant in inner space
\beqq \label{17}
{\cal H} &=& 
\frac{1}{2\,\La^2}\, \pa_{\sl 3}A_{\sl 0}\,^\al \cdot \pa_{\sl 3}A_{{\sl 0}\al} 
+ \frac{1}{2}\, \sum_{i={\sl 1}}^{\sl 2} \Pit_i\,\!^\al \cdot \Pit_{i\al} \nonumber \\
&+& \frac{1}{4\,\La^2}\, \sum_{i,j={\sl 1}}^{\sl 2} F_{ij}\,^\al \cdot F_{ij\,\al}
+ \frac{1}{2\,\La^2}\, \sum_{i={\sl 1}}^{\sl 2} \pa_{\sl 3}A_i\,^\al \cdot \pa_{\sl 3}A_{i\al} \\
&\geq& 0 \nonumber,
\eeqq
where 
\beq \label{18}
F_{ij}\,^\al \equiv \pa_i A_j\,^\al - \pa_j A_i\,^\al + A_i\,^\be \cdot \nabla_\be A_j\,^\al - A_j\,^\be \cdot \nabla_\be A_i\,^\al
\eeq
and where the additional fields $A_{\sl 0}\,^\al$ are non-local functionals of $A_i\,^\al$ and $\Pit_j\,\!^\be$ given by
\beq  \label{19}
A_{\sl 0}\,^\al \equiv \frac{1}{\pa_{\sl 3}\,\!^2} \, \La\, \sum_{i={\sl 1}}^{\sl 2} \left(\pa_i \Pit_i\,\!^\al + A_i\,^\be\cdot \nabla_\be \Pit_i\,\!^\al - \Pit_i\,\!^\be \cdot \nabla_\be A_i\,^\al \right).
\eeq
The $A_{\sl 0}\,^\al$ fulfil the unimodularity constraint
\beq \label{20}
\nabla_\al A_{\sl 0}\,^\al = 0 \eeq
which is easily proven using Eqns.(\ref{15}) and (\ref{16}).

Inverse Fourier-transforming and rotating $A_{\sl 0}\,^\al$ and $F_{ij}\,^\al$ in inner space gives the explicit non-local functional forms of ${\tilde{\hat A}}_{{\sl 0}\,a} \left[{\tilde{\hat A}}_{k\,c},\, \Pitch_{h\,d}\right], {\tilde{\hat F}}_{ij\,a} \left[{\tilde{\hat A}}_{k\,c}\right]$ appearing in ${\cal H}$ of Eqn.(\ref{1}) in terms of the original canonical variables we have used above.

Adding a trivial integration over $A_i\,^{\sl 0}$ and its conjugate field by means of including delta functions in the gauge field measures which enforce the unimodularity conditions Eqns.(\ref{15}) and (\ref{16}) on $A_i\,^\al$ and $\Pit_j\,^\be$ the gauge and "matter" field measures become 
\beqq \label{21}
& & \Pi_{\!\!\!\!\!\!_{_{_{x,X;m}}}}\!\!\!\!d\psi_m \cdot
\Pi_{\!\!\!\!\!\!_{_{_{x,X;i={\sl 1},{\sl 2},\al}}}} \!\!\!\!\!\!\!\!\!\!\!\!\!\!dA_i\,^\al \;
\Pi_{\!\!\!\!\!_{_{_{i={\sl 1},{\sl 2}}}}} \!\!\de
(\nabla_\al A_i\,^\al) \nonumber \\
& & \cdot \Pi_{\!\!\!\!\!\!_{_{_{x,X;n}}}}\!\!\!\!d\pi_n \cdot \Pi_{\!\!\!\!\!\!_{_{_{x,X;j={\sl 1},{\sl 2},\be}}}}
\!\!\!\!\!\!\!\!\!\!\!\!\!\!d\Pit_j\,^\be \;
\Pi_{\!\!\!\!\!_{_{_{j={\sl 1},{\sl 2}}}}} \!\!\de (\nabla^\be \Pit_{j\,\be})
\eeqq
with the weight
\beq \label{22}
\exp\,i\,\int\left\{\frac{1}{\La}\, \sum_{i={\sl 1}}^{\sl 2} \Pit_{i\,\al}\cdot
\pa_{\sl 0} A_i\,^\al - {\cal H} + \sum_n \pi_n\cdot \pa_{\sl 0} \psi_n - {\cal H}_M \right\}
\eeq 
which is a Lorentz-invariant expression in inner space. Note that the $\de$-functions in the integration measures above do ensure that we integrate over gauge fields and their conjugates belonging to the gauge algebra ${\overline{\bf diff}}\,{\bf M}^{\sl 4}$ only.

To keep the formulae below simple we next introduce the covariant derivative
\beq \label{23}
{\cal D}_i^\al\,\!_\be \equiv \pa_i \,\de^\al\,\!_\be + A_i\,^\ga\cdot \nabla_\ga \,\de^\al\,\!_\be - \nabla_\be A_i\,^\al
\eeq
as in \cite{chw1} allowing us e.g. to re-express
\beq  \label{24}
A_{\sl 0}\,^\al = \frac{1}{\pa_{\sl 3}\,\!^2}\, \La\, \sum_{i={\sl 1}}^{\sl 2} {\cal D}_i^\al\,_\be \,\Pit_i\,\!^\be
\eeq
in a compact form.

The third step is to turn the path integrals above into Lorentz-invariant expressions as well in space-time applying the usual trick to treat $A_{\sl 0}\,^\al$ as a new independent variable which we can integrate over \cite{stw2}. The trick still works with the constrained measure Eqn.(\ref{21}). In fact, as the weight factor Eqn.(\ref{22}) is at most quadratic in $A_{\sl 0}\,^\al$ we find that
\beqq \label{25}
& & \quad \int \Pi_{\!\!\!\!\!\!_{_{_{x,X;\al}}}}
\!\!\!\!dA_{\sl 0}\,^\al \;
\, \de (\nabla_\al A_{\sl 0}\,^\al) \nonumber \\
& & \cdot \exp\,i\,\int\left\{\frac{1}{\La}\, \sum_{i={\sl 1}}^{\sl 2} \Pit_{i\al}\cdot
\pa_{\sl 0} A_i\,^\al - {\cal H} \right\} \nonumber \\
& & \propto \int \Pi_{\!\!\!\!\!\!_{_{_{x,X;\al}}}}
\!\!\!\!d\TA_{\sl 0}\,^\al \;
\, \de (\nabla_\al \TA_{\sl 0}\,^\al) \\
& & \cdot \exp -\frac{i}{2}\,\int \, \frac{1}{\La^2}\, \TA_{\sl 0}\,^\al \cdot
\pa_{\sl 3}\,\!\!^2 \TA_{\sl 0\,\al } \nonumber \\
& & \cdot \exp \,i\,\int \frac{1}{2}\, \sum_{i,j={\sl 1}}^{\sl 2} 
{\cal D}_i^\al\,_\be\, \Pit_i\,\!^\be \cdot \frac{1}{\pa_{\sl 3}\,\!^2}\,
{\cal D}_{j\al}\,^\ga\, \Pit_{j\ga} +\dots \nonumber \\
& & \propto \exp\,i\,\int \left\{ \frac{1}{\La}\, \sum_{i={\sl 1}}^{\sl 2} \Pit_{i\al}\cdot \pa_{\sl 0} A_i\,^\al - {\cal H} \right\}
\nonumber \eeqq
after a shift of integration variables $\,\TA_{\sl 0}\,^\al \equiv A_{\sl 0}\,^\al - \frac{1}{\pa_{\sl 3}\,\!^2}\, \La \sum_{i={\sl 1}}^{\sl 2} {\cal D}_i^\al\,\!_\be\, \Pit_i\,\!^\be$. Apart from a field-independent normalization factor this is the gauge weight factor Eqn.(\ref{22}) with $A_{\sl 0}\,^\al$ given by Eqn.(\ref{24}) in terms of $A_i\,^\al$, $\Pit_j\,\!^\be$.

The fourth step is to perform the corresponding $\Pit_j\,\!^\be$ integrations for fixed $A_i\,^\al$ and $A_{\sl 0}\,^\al$ which is possible because ${\cal H}$ is quadratic in $\Pit_j\,\!^\be$. After a shift of integration variables $\:\TPit_j\,^\al \equiv \Pit_j\,^\al - \frac{1}{\La}\, F_{{\sl 0}j}\,^\al$ we find
\beqq \label{26}
& & \quad \int \Pi_{\!\!\!\!\!\!_{_{_{x,X;j={\sl 1},{\sl 2},\al}}}}
\!\!\!\!\!\!\!\!\!\!\!\!\!\!d\, \Pit_j\,^\al \;
\Pi_{\!\!\!\!\!_{_{_{j={\sl 1},{\sl 2}}}}} \!\!\de (\nabla^\al \Pit_{j\,\al}) \nonumber \\
& & \cdot \exp\,i\,\int \left\{-\frac{1}{2}\, \sum_{i={\sl 1}}^{\sl 2} \Pit_i\,\!^\al \cdot \Pit_{i\,\al} + \frac{1}{\La}\,\sum_{i={\sl 1}}^{\sl 2} F_{{\sl 0}i}\,^\al \cdot \Pit_{i\,\al} 
+\dots \right\} \nonumber \\
& & \propto \int \Pi_{\!\!\!\!\!\!_{_{_{x,X;j={\sl 1},{\sl 2},\al}}}}
\!\!\!\!\!\!\!\!\!\!\!\!\!\!d\, \TPit_j\,^\al \;
\Pi_{\!\!\!\!\!_{_{_{j={\sl 1},{\sl 2}}}}} \!\!\de (\nabla^\al \TPit_{j\,\al}) \\
& & \cdot \exp\,i\,\int \left\{-\frac{1}{2}\, \sum_{i={\sl 1}}^{\sl 2} \TPit_i\,\!^\al \cdot \TPit_{i\,\al} + \int \frac{1}{2\, \La^2}\, \sum_{i={\sl 1}}^{\sl 2} F_{{\sl 0}i}\,^\al \cdot F_{{\sl 0}i\,\al}
+\dots\right\}  \nonumber \\
& & \propto \exp \,i\,\int \frac{1}{2\, \La^2}\, \sum_{i={\sl 1}}^{\sl 2}
F_{{\sl 0}i}\,^\al \cdot F_{{\sl 0}i\,\al} +\dots\:.
\nonumber \eeqq
Above we have introduced
\beq \label{27}
F_{{\sl 0}i}\,^\al \equiv \pa_{\sl 0} A_i \,^\al - \pa_i A_{\sl 0}\,^\al + A_{\sl 0}\,^\be \cdot \nabla_\be A_i\,^\al - A_i\,^\be \cdot \nabla_\be A_{\sl 0}\,^\al 
\eeq
and used that $ F_{{\sl 0}i}\,^\al$ is an element of the gauge algebra ${\overline{\bf diff}}\,{\bf M}^{\sl 4}$ as is easily verified.

As a result Green functions are given as path integrals over $A_i\,^\al$, $A_{\sl 0}\,^\al$ and $\psi_m$ - assuming that the integrations over $\pi_n$ are Gaussian as well - with the gauge field measure 
\beq \label{28}
\Pi_{\!\!\!\!\!\!_{_{_{x,X;\al}}}}
\!\!\!\!dA_{\sl 0}\,^\al \;
\, \de (\nabla_\al A_{\sl 0}\,^\al)
\cdot \Pi_{\!\!\!\!\!\!_{_{_{x,X;i={\sl 1},{\sl 2},\al}}}} \!\!\!\!\!\!\!\!\!\!\!\!\!\!dA_i\,^\al \;
\Pi_{\!\!\!\!\!\!_{_{_{i={\sl 1},{\sl 2}}}}} \!\!\de (\nabla_\al A_i\,^\al)
\eeq
and gauge field weight
\beqq \label{29}
& &\!\!\!\!\!\!\!\!\exp\,i\,\int\Bigg\{ \frac{1}{2\, \La^2}\,\sum_{i={\sl 1}}^{\sl 2} F_{{\sl 0}i}\,^\al \cdot F_{{\sl 0}i\,\al}
- \frac{1}{4\, \La^2}\, \sum_{i,j={\sl 1}}^{\sl 2} F_{ij}\,^\al \cdot F_{ij\,\al} \nonumber \\
& & - \frac{1}{2\, \La^2}\, \sum_{i={\sl 1}}^{\sl 2} \pa_{\sl 3}A_i\,^\al\cdot 
\pa_{\sl 3}A_{i\,\al} + \frac{1}{2\, \La^2}\, \pa_{\sl 3}A_{\sl 0}\,^\al\cdot 
\pa_{\sl 3}A_{{\sl 0}\,\al} \Bigg\}.
\eeqq

The last step is to introduce the variable $A_{\sl 3}\,^\al$ which vanishes identically in the Minkowski-plus-axial gauge and to finally recast the path integrals in a manifestly Lorentz- and gauge-invariant fashion in both space-time and inner space with gauge field measure
\beq \label{30}
\Pi_{\!\!\!\!\!\!_{_{_{x,X;\mu,\al}}}}
\!\!\!\!\!\!\!\!dA_\mu\,^\al \;
\Pi_{\!\!\!\!\!_{_{_{\mu}}}} \,\, \de (\nabla_\al A_\mu\,^\al) 
\eeq
and gauge field weight
\beq \label{31}
\de (A_{\sl 3}\,^\al)\cdot \exp\,i\,\int \Big\{{\cal L} + \ep \mbox{-terms} \Big\},
\eeq
where
\beq \label{32}
{\cal L} = - \frac{1}{4\, \La^2}\, F_{\mu\nu}\,^\al \cdot F^{\mu\nu}\,_\al \eeq
is the Lagrangian density of the gauge theory of volume-preserving diffeomorphisms of ${\bf M}^{\sl 4}$ and
\beq \label{33}
F_{\mu\nu}\,^\al = \pa_\mu A_\nu\,^\al - \pa_\nu A_\mu\,^\al + A_\mu\,^\be \cdot \nabla_\be A_\nu\,^\al - A_\nu\,^\be \cdot \nabla_\be A_\mu\,^\al \eeq
are the covariant field strength components \cite{chw1}. The $\ep$-terms indicate the appropriate imaginary parts of propagators.

Note that the measure Eqn.(\ref{30}) is the gauge-invariant functional measure on the space of gauge fields living in the gauge algebra ${\overline{\bf diff}}\,{\bf M}^{\sl 4}$. Also note that the path integrals are to be evaluated over functional spaces which are defined such that their Fourier-transformed elements have support on ${\bf V^+}(K)\cup {\bf V^-}(K)$.

\section{General Gauge Fixing in the De Witt-Faddeev-Popov Approach and Ghosts}
In this section we define the quantum gauge field theory of volume-preserving diffeomorphisms of ${\bf M}^{\sl 4}$ in general gauges based on the De Witt-Faddeev-Popov (FP) method introducing the ghost fields related to these gauges and the generating functional for Green functions.

Following closely \cite{stw2} we start noting that gauge-invariant Green functions calculated as path integrals with measure and weight given by Eqns. (\ref{30}) and (\ref{31}) respectively are of the general form
\beq \label{34}
{\cal J} = \int\, \Pi_{\!\!\!\!\!\!_{_{_{x,X;n}}}}\!\!\!\!d \phi_n \cdot
{\cal G} \left[\phi \right] B \left[ f[\phi] \right]
\Det {\cal F} \left[\phi \right],
\eeq
where $\phi_n (x,X)$ are a set of gauge and matter fields, $\Pi_{\!\!\!\!\!\!_{_{_{x,X;n}}}}\!\!\!\!d \phi_n$ is a volume element and ${\cal G} \left[\phi \right]$ is a functional of the $\phi_n$ satisfying the gauge-invariance requirement
\beq \label{35}
\Pi_{\!\!\!\!\!\!_{_{_{x,X;n}}}}\!\!\!\!d \phi_{_{\cal E}\,n} \cdot
{\cal G} \left[\phi_{_{\cal E}} \right] =^{\!\!\!\!{!}}\:
\Pi_{\!\!\!\!\!\!_{_{_{x,X;n}}}}\!\!\!\!d \phi_n \cdot
{\cal G} \left[\phi \right].
\eeq
$\phi_{_{\cal E}\,n}$ denote the fields after an infinitesimal gauge transformation with local gauge parameters ${\cal E}^\al (x,X)$ \cite{chw1}, $f_\ga [\phi;x,X]$ is a vector-valued non gauge-invariant gauge-fixing functional, $B \left[ f \right]$ a numerical functional defined on general $f$ and ${\cal F}$ is the operator
\beq \label{36}
{\cal F}^\ga\,_\de \left[\phi \right](x,X) \equiv \frac{\de f^\ga [\phi_{_{\cal E}}](x,X)}{\de\, {\cal E}^\de (x,X)} _{\mid_{_{{\cal E}=0}}}. \eeq
Indeed, with fields $\phi_n$ taken as $A_\mu\,^\al$ and $\psi_m$, and setting
\beqq \label{37}
f^\ga [A,\psi] &=& A_{\sl 3}\,\!^\ga, \nonumber \\
B \left[ f \right] &=& \Pi_{\!\!\!\!\!\!_{_{_{x,X;\ga}}}}
\!\!\!\de\left(f^\ga (x,X) \right), \nonumber \\
{\cal G} [A,\psi] &=& \exp\,i\,\int \Big\{{\cal L} + {\cal L}_M +  \ep \mbox{-terms} \Big\} \\
&\times & \mbox{gauge-invariant functionals of $A,\psi$} \nonumber \\
\Pi_{\!\!\!\!\!\!_{_{_{x,X;n}}}}\!\!\!\!d \phi_n &=&
\Pi_{\!\!\!\!\!\!_{_{_{x,X;m}}}}
\!\!\!\!d\psi_m \cdot
\Pi_{\!\!\!\!\!\!_{_{_{x,X;\mu,\al}}}}
\!\!\!\!\!\!\!\!dA_\mu\,^\al \;
\Pi_{\!\!\!\!\!_{_{_{\mu}}}} \,\, \de (\nabla_\al A_\mu\,^\al)
\nonumber \eeqq
the integral ${\cal J}$ Eqn.(\ref{34}) yields the Green functions of the gauge theory of volume-preserving diffeomorphisms of ${\bf M}^{\sl 4}$ in the Minkowski-plus-axial gauge as defined above. Here we have used the fact that
\beq \label{38} {\cal F}^\ga\,_\de [A_\mu\,^\al](x,X) =
\eta^\ga\,_\de\cdot \pa_{\sl 3} \eeq
is field-independent and $\Det {\cal F}$ reduces to an overall normalization factor in the Minkowski-plus-axial gauge.

Next, let us check the gauge-invariance requirement Eqn.(\ref{35}). Under local gauge transformations we have \cite{chw1}
\beqq \label{39}
& & A_{_{\cal E}\,\mu}\,^\al = A_\mu\,^\al + \pa_\mu {\cal E}^\al + A_\mu\,^\be \cdot \nabla_\be {\cal E}^\al - {\cal E}^\be \cdot \nabla_\be A_\mu\,^\al, \nonumber \\
& & \Pi_{\!\!\!\!\!\!_{_{_{x,X;\mu,\al}}}}\!\!\!\!\!\!\!\!dA_{_{\cal E}\,\mu}\,^\al = \Det \left( \frac{\de A_{_{\cal E}\,\mu}\,^\al}{\de A_\nu\,^\be} \right) \cdot \Pi_{\!\!\!\!\!\!_{_{_{x,X;\mu,\al}}}}\!\!\!\!\!\!\!\!dA_\mu\,^\al, \\
& & \de (\nabla_\al A_{_{\cal E}\,\mu}\,^\al) = \de (\nabla_\al A_\mu\,^\al). \nonumber
\eeqq
Calculating
\beq \label{40}
\frac{\de A_{_{\cal E}\,\mu}\,^\al}{\de A_\nu\,^\be} = \eta_\mu\,^\nu\cdot\left( \eta^\al\,_\be + \nabla_\be {\cal E}^\al - {\cal E}^\ga\cdot
\nabla_\ga\, \de^\al \,_\be \right) 
\eeq
we find that the functional trace of the logarithm of the above Jacobian vanishes - yielding $\Det (\dots)=1$ in Eqn.(\ref{39}). As a result the gauge field measure is gauge-invariant and the condition Eqn.(\ref{35}) is fulfilled.

Now we are in a position to freely change the gauge as path integrals of the form Eqn.(\ref{34}) are actually independent of the gauge-fixing functional $f^\ga [\phi;x,X]$ and depend on the choice of the functional $B \left[ f \right]$ only through an irrelevant constant. The proof of this crucial theorem is found e.g. in \cite{stw2} - as all the steps in the proof hold true for the gauge theory of volume-preserving diffeomorphisms of ${\bf M}^{\sl 4}$ as well we do not repeat them explicitly here.

As a result the generating functional for the Green functions of the gauge theory of volume-preserving diffeomorphisms of ${\bf M}^{\sl 4}$ in an arbitrary gauge and in the presence of "matter" fields is given by
\beqq \label{41} {\cal Z}\left[\eta, J \right]
&\equiv& \int\Pi_{\!\!\!\!\!\!_{_{_{x,X;m}}}}
\!\!\!\!d\psi_m \cdot
\int\Pi_{\!\!\!\!\!\!_{_{_{x,X;\mu,\al}}}}
\!\!\!\!\!\!\!\!dA_\mu\,^\al \;
\Pi_{\!\!\!\!\!_{_{_{\mu}}}} \,\,
\de (\nabla_\al A_\mu\,^\al) \\
& & \cdot \exp\,i\,\left\{S + S_M + \frac{1}{\La^2}\, \int J\cdot A + \int \sum_m \eta_m\cdot \psi_m + \ep \mbox{-terms} \right\} \nonumber \\
& & \cdot B \left[ f [A,\psi] \right]
\Det {\cal F} \left[A,\psi \right], \nonumber \eeqq
where we have introduced the external sources $\eta$ and $J$ - transforming as a vector in inner space - for the "matter" and gauge fields respectively.

In order to further evaluate the generating functional above we choose
\beqq \label{42}
B\left[f [A,\psi] \right]&\equiv& \exp \,i \, S_{GF} \nonumber \\
S_{GF} &\equiv& - \frac{1}{2\xi\, \La^2}\,\intx\intX \La^{-4} \,
f_\ga [A,\psi] \cdot f^\ga [A,\psi]
\eeqq
to be quadratic in the gauge-fixing functional $f^\ga [A,\psi]$ which transforms as a vector in inner space and re-express the functional determinant as the Gaussian integral
\beqq \label{43}
\Det {\cal F} \left[A,\psi \right] &\propto&
\int\,\Pi_{\!\!\!\!\!\!_{_{_{x,X;\ga}}}}\!\!d\om^*_\ga \;
\de (\nabla^\ga \om^*_\ga) \cdot \int\,\Pi_{\!\!\!\!\!\!_{_{_{x,X;\de}}}}\!\!d\om^\de \;
\de (\nabla_\de\, \om^\de) \cdot \exp \,i \, S_{GH} \nonumber \\
S_{GH} &\equiv& \frac{1}{\La^2} \, \intx\intX \La^{-4} \, \om^*_\ga \cdot
{\cal F}^\ga\,_\de \left[A,\psi \right] \om^\de.
\eeqq
Above we have introduced the ghost fields $\om^*_\ga (x,X)$ and $\om^\de (x,X)$ which are independent anti-commuting classical variables. The $\de$-functions ensure that both sets of variables obey the same constraints as the gauge parameters ${\cal E}$ and that the corresponding operators $\om^* \equiv \om^*_\ga \nabla^\ga$ and $\om \equiv \om^\de \nabla_\de$ are elements of the gauge algebra ${\overline{\bf diff}}\,{\bf M}^{\sl 4}$ which proves crucial in defining the BRST-symmetry operation later.

What is the condition to represent $\Det {\cal F} \left[A,\psi \right]$ above as a Gaussian integral as in Eqn.(\ref{43})?

The condition is that for $\om^\de$ in the gauge algebra ${\cal F}^\ga\,_\de\, \om^\de$ is in the gauge algebra as well. Then
\beq \label{44}
{\cal F}^\ga\,_\de:\: {\overline{\bf diff}}\,{\bf M}^{\sl 4}\: \lar \:\: {\overline{\bf diff}}\,{\bf M}^{\sl 4}
\eeq
is an endomorphism of ${\overline{\bf diff}}\,{\bf M}^{\sl 4}$.

Defining the scalar product
\beq \label{45}
\langle g \!\mid\! h \rangle \equiv \frac{1}{\La^2} \, \intX \La^{-4}\, g_\al^\dagger (x,X) \cdot h^\al (x,X)
\eeq
on ${\overline{\bf diff}}\,{\bf M}^{\sl 4}$ and restricting ourselves to vector-valued functions in ${\overline{\bf diff}}\,{\bf M}^{\sl 4}$ which are square-integrable in the sense of the scalar product above the corresponding function space becomes a Hilbert space. For ${\cal F}^\ga\,_\de$ being a selfadjoint endomorphism of ${\overline{\bf diff}}\,{\bf M}^{\sl 4}$ with a complete system of orthonormal eigenvectors we indeed have Eqn.(\ref{43}) with the $\de$-functions automatically taken account of in the Gaussian integration.

Finally we can write the generating functional for the Green functions of the gauge theory of volume-preserving diffeomorphisms of ${\bf M}^{\sl 4}$ in an arbitrary gauge as
\beqq \label{46}
{\cal Z}\left[\eta, J \right]
&=& \int\Pi_{\!\!\!\!\!\!_{_{_{x,X;m}}}}
\!\!\!\!d\psi_m \cdot
\int\Pi_{\!\!\!\!\!\!_{_{_{x,X;\mu,\al}}}}
\!\!\!\!\!\!\!\!dA_\mu\,^\al \;
\Pi_{\!\!\!\!\!_{_{_{\mu}}}} \,\,
\de (\nabla_\al A_\mu\,^\al) \nonumber \\
& & \!\!\!\!\!\!\!\!\!\!\!\!
\cdot \int\,\Pi_{\!\!\!\!\!\!_{_{_{x,X;\ga}}}}\!\!d\om^*_\ga \;
\de (\nabla^\ga \om^*_\ga)
\cdot \int\,\Pi_{\!\!\!\!\!\!_{_{_{x,X;\de}}}}\!\!d\om^\de \;
\de (\nabla_\de\, \om^\de) \\
& & \!\!\!\!\!\!\!\!\!\!\!\!
\cdot \exp\,i\,\left\{S_{MOD} + S_M + \frac{1}{\La^2}\, \int J\cdot A + \int \sum_m \eta_m\cdot \psi_m + \ep \mbox{-terms} \right\}, \nonumber  \eeqq
where
\beq \label{47}
S_{MOD}\equiv S + S_{GF} + S_{GH} \eeq
is the modified FP gauge-fixed action for the gauge theory of volume-preserving diffeomorphisms of ${\bf M}^{\sl 4}$  .

Eqn.(\ref{46}) defines the quantum gauge field theory of volume-preserving diffeomorphisms of ${\bf M}^{\sl 4}$ and is the starting point for the evaluation of matrix elements at the quantum level.

\section{Perturbative Expansion, Feynman Rules and Asymptotic States}
In this section we derive the perturbative expansion of the generating functional for the Green functions of the quantum gauge field theory of volume-preserving diffeomorphisms of ${\bf M}^{\sl 4}$ and its Feynman rules in Lorentz-covariant Minkowski gauges. We then use power counting to demonstrate the superficial renormalizability of the quantum gauge field theory of volume-preserving diffeomorphisms of ${\bf M}^{\sl 4}$. Finally we analyze the asymptotic states of the theory and are led to introduce additional quantum numbers related to the inner degrees of freedom of the theory.

Working in Minkowski gauges with inner metric $\eta_{\al\be}$ we use Eqn.(\ref{46}) as the starting point for perturbation theory. Splitting the action
\beq \label{48}
S_{MOD} [A, \om^*, \om] \equiv S_{\sl 0} [A, \om^*, \om]
+ S_{INT} [A, \om^*, \om] \eeq
into the part $S_{\sl 0}$ quadratic in the gauge and ghost fields and the interaction part $S_{INT}$ we can rewrite Eqn.(\ref{46}) for the pure gauge field theory as
\beq \label{49}
{\cal Z}\left[J, \ze^*, \ze \right] = \exp\,i\, S_{INT} \left[\frac{\de\rvec }{\de J}, \frac{\de\rvec}{\de \ze} ,\frac{\de\rvec}{\de \ze^*}\right]  \, {\cal Z}_{\sl 0}\left[J, \ze^*, \ze \right],
\eeq
where
\beqq \label{50}
{\cal Z}_{\sl 0}\left[J, \ze^*, \ze \right]
&\equiv&
\int\Pi_{\!\!\!\!\!\!_{_{_{x,X;\mu,\al}}}}
\!\!\!\!\!\!\!\!dA_\mu\,^\al \;
\Pi_{\!\!\!\!\!_{_{_{\mu}}}} \,\,
\de (\nabla_\al A_\mu\,^\al) \nonumber \\
& & \cdot \int\,\Pi_{\!\!\!\!\!\!_{_{_{x,X;\ga}}}}\!\!d\om^*_\ga \;
\de (\nabla^\ga \om^*_\ga)
\cdot \int\,\Pi_{\!\!\!\!\!\!_{_{_{x,X;\de}}}}\!\!d\om^\de \;
\de (\nabla_\de\, \om^\de) \\
& & \cdot \exp\,i\,\left\{S_{\sl 0} + \frac{1}{\La^2}\, \int (J\cdot A + \om^*\cdot \ze + \ze^*\cdot \om) + \ep \mbox{-terms} \right\} \nonumber \eeqq
is the generating functional for Green functions of the non-interacting theory and $\ze$, $\ze^*$ are sources for the ghost fields. Note that for consistency reasons all $J$, $\ze$, $\ze^*$ have to be elements of the gauge algebra ${\overline{\bf diff}}\,{\bf M}^{\sl 4}$. In particular, it is crucial that the conserved gauge-field currents
\beq \label{51} J_\nu\,^\be = A^{\mu\,\al}\cdot \nabla_\al F_{\mu\nu}\,^\be
- F_{\mu\nu}\,^\al\cdot \nabla_\al A^{\mu\,\be}
\eeq
related to the global coordinate transformation invariance in inner space and generating the self coupling of the gauge fields are elements of the gauge algebra ${\overline{\bf diff}}\,{\bf M}^{\sl 4}$ which is easily verified.

To derive Feynman rules we have to specify the gauge and choose
\beq \label{52}
f^\ga [A] \equiv \pa^\mu A_\mu\,^\ga
\eeq
as the Lorentz-covariant gauge fixing function resulting in
\beq \label{53}
{\cal F}^\ga\,_\de \left[A\right] = \pa^\mu\! \left(\pa_\mu \,\eta^\ga\,_\de + A_\mu\,^\al\cdot \nabla_\al \,\eta^\ga\,_\de - \nabla_\de A_\mu\,^\ga\right) 
\eeq
which is easily shown to be an endomorphism of ${\overline{\bf diff}}\,{\bf M}^{\sl 4}$ as required.

For the choice Eqn.(\ref{52}) $S_{\sl 0}$ is calculated to be
\beqq \label{54}
S_{\sl 0} &=& - \frac{1}{2\, \La^2} \, \int \, A_\mu\,^\al \cdot
{\cal D}_{{\sl 0},\xi}^{\mu\nu}\,_{\al\be}\, A_\nu\,^\be \nonumber \\
&-& \frac{1}{\La^2} \, \int \, \om^*_\ga \cdot {\cal D}_{\sl 0}^\ga\,_\de\, \om^\de,
\eeqq
where we have defined the non-interacting gauge and ghost field fluctuation operators by
\beqq \label{55}
{\cal D}_{{\sl 0},\xi}^{\mu\nu}\,_{\al\be} &\equiv& \left( -\, \eta ^{\mu\nu} \cdot \pa^2  + \left(1 - \frac{1}{\xi} \right)\,
\pa^\mu \, \pa^\nu \right) \eta_{\al\be} \nonumber \\
{\cal D}_{\sl 0}^\ga\,_\de &\equiv& -\, \pa^2\, \eta^\ga\,_\de
\eeqq
and the corresponding free propagators $G^{\sl 0}$ through
\beqq \label{56}
{\cal D}_{{\sl 0},\xi}^{\mu\rho}\,_{\al\ga} \,
G^{{\sl 0},\xi}_{\rho\nu}\,^{\ga\be} (x,y; X,Y) &=& \La^4\,\, ^T\!\de_\al\,^\be (X-Y) \, \eta^\mu\,_\nu \, \de^4 (x-y) \nonumber \\
{\cal D}_{\sl 0}^\ga\,_\al \, G_{\sl 0}^\al\,_\de (x,y; X,Y) &=& \La^4\,\, ^T\!\de^\ga\,_\de (X-Y)  \, \de^4 (x-y),
\eeqq
where
\beq \label{}
^T\!\de_{\al\be}(X-Y) = \int\! \frac{d^4 K}{(2{\pi})^4}\, \La^4\, e^{-iK(X-Y)}
\left( \eta_{\al\be} - \frac{K_\al K_\be}{K^2}\right)
\eeq 
is the delta function transversal in inner space. The factors of $\La$ ensure the scale invariance of the r.h.s under inner scale transformations.

After some algebra we find the propagators for the gauge and ghost fields to be
\beqq \label{57}
& & G^{{\sl 0},\xi}_{\mu\nu}\,^{\al\be} (x,y; X,Y) =
\La^4\,\, ^T\!\de^{\al\be} (X-Y) 
\nonumber \\
& & \quad\quad\quad\
\cdot \int\! \frac{d^4 k}{(2{\pi})^4}\, e^{i\,k\cdot (x-y)}
\, \frac{1}{k^2 - i\,\ep} \left(\eta_{\mu\nu} -
(1-\xi) \frac{k_\mu k_\nu}{k^2} \right) \\
& & G_{\sl 0}^\ga\,_\de (x,y; X,Y) = 
\La^4\,\, ^T\!\de^\ga\,_\de (X-Y)
\nonumber \\
& & \quad\quad\quad
\cdot \int\! \frac{d^4 k}{(2{\pi})^4}\, e^{i\,k\cdot (x-y)}
\, \frac{1}{k^2 - i\,\ep}. \nonumber \eeqq
They are manifestly diagonal and local in inner space and invariant under local inner Poincar\'e transformations $X^\al\ar X'^\al = T^\al (x) + \La^\al\,_\be (x)\, X^\be$, $\La^\al\,_\be \in SO(1,3)$.

Both the fluctuation operators and the propagators are endomorphisms of ${\overline{\bf diff}}\,{\bf M}^{\sl 4}$, i.e. if $f^\al$ fulfills $\nabla_\al f^\al =0$ so will ${\cal D}_{{\sl 0},\xi}^{\mu\nu}\,_{\al\be} f^\be$, $G^{{\sl 0},\xi}_{\mu\nu}\,^{\al\be} f_\be$ and ${\cal D}_{\sl 0}^\ga\,_\de f^\de$, $G_{\sl 0}^\ga\,_\de f^\de$ as is easily verified. In other words the propagators are the inverses of the fluctuation operators on the functional space ${\overline{\bf diff}}\,{\bf M}^{\sl 4}$. As a consequence the $\de$-functions in the measure in Eqn.(\ref{50}) will be automatically taken care of in the Gaussian integrals above.

Performing the Gaussian integrals over the gauge and ghost fields we find
\beqq \label{58}
{\cal Z}_{\sl 0}\left[J, \ze^*, \ze \right]
&\propto&
\exp\,i\, \frac{1}{2\, \La^2} \int\!\!\int J^\mu\,_\al\cdot
G^{{\sl 0},\xi}_{\mu\nu}\,^{\al\be} \, J^\nu\,_\be \nonumber\\
& &
\cdot \exp\,i\, \frac{1}{\La^2} \int\!\!\int \ze^*_\ga\cdot
G_{\sl 0}^\ga\,_\de \, \ze^\de
\eeqq
up to the functional determinants of the fluctuation operators Eqns.(\ref{55}). These field-independent normalization factors do not contribute to physical amplitudes and can be discarded.

Insertion of the result above into Eqn.(\ref{49}) gives the unrenormalized perturbation expansion of the generating functional of the Green functions of the quantum gauge field theory of volume-preserving diffeomorphisms of ${\bf M}^{\sl 4}$ which is plagued by the usual ultraviolet and infrared divergencies of perturbative QFT. On top of these divergencies we will have to deal with potentially divergent integrals over inner space. We will show below that they can be consistently defined respecting the inner scale invariance of the classical theory.

Next we give the momentum space Feynman rules which are easily derived generalizing the usual approach by Fourier-transforming inner space integrals as well.

The momentum space gauge field and ghost propagators are given by
\beqq \label{59}
G^{{\sl 0},\xi}_{\mu\nu}\,^{\al\be} (k; K) &=&
\frac{1}{k^2 - i\,\ep} \left(\eta_{\mu\nu} - (1-\xi) \frac{k_\mu k_\nu}{k^2} \right) \left( \eta_{\al\be} - \frac{K_\al K_\be}{K^2}\right)
\nonumber \\
G_{\sl 0}^\ga\,_\de (k; K) &=& \frac{1}{k^2 - i\,\ep}\, 
\left( \eta^\ga\,_\de - \frac{K^\ga K_\de}{K^2}\right)
\eeqq
being transversal in inner space. The inner degrees of freedom do not propagate whereas the space-time parts of the propagators equal the well-known Yang-Mills propagators.

The particle content is now easily read off - there is an uncountably infinite number of both massless gauge and unphysical ghost fields - the latter to counter-balance the unphysical gauge field degrees of freedom arising in covariant gauges. 

Next we calculate the vertices starting with the tri-linear gauge field self-coupling
\beq \label{60}
-\, \La^{-2}\, \left(\pa_\mu A_\nu\,^\al - \pa_\nu A_\mu\,^\al\right)\, A^\mu\,_\be \cdot \nabla^\be A^\nu\,_\al
\eeq
corresponding to a vertex with three vector boson lines. If these lines carry incoming space-time momenta $k_1$, $k_2$, $k_3$, inner momentum space coordinates $K_1$, $K_2$, $K_3$ and gauge field indices $\mu \al$, $\nu \be$, $\rho \ga$ the contribution of such a vertex to a Feynman integral is
\beqq \label{61}
& & -\, 2\, \La^{-2}\, \Big\{ K_1^\ga\,\eta^{\al\be}\,( k_{2\,\rho} \eta_{\mu\nu} - k_{2\,\mu} \eta_{\nu\rho}) \nonumber \\
& &\quad\quad +\,\, K_2^\al\,\eta^{\be\ga}\,( k_{3\,\mu} \eta_{\nu\rho} - k_{3\,\nu} \eta_{\rho\mu}) \\
& &\quad\quad +\,\, K_3^\be\,\eta^{\ga\al}\,( k_{1\,\nu} \eta_{\rho\mu} - k_{1\,\rho} \eta_{\mu\nu}) \Big\} \nonumber 
\eeqq
with
\beq \label{62}
k_1+k_2+k_3=0,\quad\quad K_1+K_2+K_3=0.
\eeq

The quadri-linear gauge field self-coupling term
\beq \label{63}
-\, \frac{1}{2\, \La^2}\, \left(A_\mu\,^\be \cdot \nabla_\be A_\nu\,^\al - A_\nu\,^\be \cdot \nabla_\be A_\mu\,^\al\right)\, A^\mu\,_\ga \cdot \nabla^\ga A^\nu\,_\al
\eeq
corresponds to a vertex with four vector boson lines. If these lines carry incoming space-time momenta $k_1$, $k_2$, $k_3$, $k_4$, inner momentum space coordinates $K_1$, $K_2$, $K_3$, $K_4$ and gauge field indices $\mu \al$, $\nu \be$, $\rho \ga$, $\si \de$ the contribution of such a vertex to a Feynman integral is
\beqq \label{64}
& & -\, \La^{-2}\, \Big\{( K_1^\ga\, K_2^\de\, \eta^{\al\be} - K_2^\de\, K_3^\al\, \eta^{\be\ga}
+ K_3^\al\, K_4^\be\, \eta^{\ga\de} - K_1^\ga\, K_4^\be\, \eta^{\al\de})
\nonumber \\
& & \quad\quad\quad\quad\quad\quad\quad\quad\quad\quad\quad \cdot( \eta_{\mu\nu}\eta_{\rho\si} - \eta_{\mu\si}\eta_{\nu\rho}) \nonumber \\
& & \quad\quad +\,\, ( K_1^\de\, K_2^\ga\, \eta^{\al\be} - K_1^\de\, K_3^\be\, \eta^{\al\ga}
+ K_3^\be\, K_4^\al\, \eta^{\ga\de} - K_2^\ga\, K_4^\al\, \eta^{\be\de})
\nonumber \\
& & \quad\quad\quad\quad\quad\quad\quad\quad\quad\quad\quad \cdot( \eta_{\mu\nu}\eta_{\rho\si} - \eta_{\mu\rho}\eta_{\nu\si}) \\
& & \quad\quad +\,\, ( K_1^\be\, K_3^\de\, \eta^{\al\ga} - K_1^\be\, K_4^\ga\, \eta^{\al\de}
+ K_2^\al\, K_4^\ga\, \eta^{\be\de} - K_2^\al\, K_3^\de\, \eta^{\be\ga})
\nonumber \\
& & \quad\quad\quad\quad\quad\quad\quad\quad\quad\quad\quad \cdot( \eta_{\mu\rho}\eta_{\nu\si} - \eta_{\mu\si}\eta_{\nu\rho})\Big\} \nonumber
\eeqq
with
\beq \label{65}
k_1+k_2+k_3+k_4 = 0,\quad\quad K_1+K_2+K_3+K_4 = 0.
\eeq

Finally, the gauge-ghost field coupling term
\beq \label{66}
-\, \La^{-2}\, \pa^\mu \om^*_\ga \left(A_\mu\,^\de\cdot \nabla_\de \,\om^\ga -\, \om^\de\cdot \nabla_\de A_\mu\,^\ga\right)
\eeq
corresponds to a vertex with one outgoing and one incoming ghost line as well as one vector boson line. If these lines carry incoming space-time momenta $k_1$, $k_2$, $k_3$, inner momentum space coordinates $K_1$, $K_2$, $K_3$ and field indices $\ga$, $\de$, $\mu \al$ the contribution of such a vertex to a Feynman integral becomes
\beq \label{67}
-\, \La^{-2}\, ( K_2^\al\,\eta^{\ga\de} -\, K_3^\de\,\eta^{\al\be})\, k_{1\,\mu}
\eeq
with
\beq \label{68}
k_1+k_2+k_3=0,\quad\quad K_1+K_2+K_3=0.
\eeq

In summary, the above propagators and vertices allow us to perturbatively evaluate the Green functions of the theory. Note that for any Feynman graph the analogon of the sums over Lie algebra structure constants in Yang-Mills theories are integrals over inner momentum space variables with the scale-invariant measure
\beq \label{69}
\int\! \frac{d^4 K}{(2{\pi})^4}\,\La^4.
\eeq
As the vertices in such graphs contribute polynomials in the inner space coordinates $K_\al$ to the integrand and as these inner degrees of freedom do not propagate such integrals look badly divergent - we will show in the next section that they can be {\it consistently defined} respecting the inner scale invariance of the classical theory.

Turning to the space-time integrals and renormalizability in the power-counting sense we note that the gauge and ghost fields have the same canonical dimensions $[A]=1$ and $[\om^*]=[\om]=1$ relevant for power counting as their Yang-Mills counterparts do. 

The corresponding divergence indices $\de_1$ of the tri-linear gauge field vertex, $\de_2$ of the quadri-linear gauge field vertex and $\de_3$ of the ghost-gauge field vertex vanish
\beq \label{70}
\de_1 = \de_3 = b + d - 4 = 3 + 1 - 4 = 0,\quad\quad
\de_2 = 4 - 4 = 0,
\eeq
where $b$ is the number of gauge field and ghost lines and $d$ the number of space-time derivatives attached to the respective vertex.

Accordingly the superficial degree of divergence $\om$ for any diagram with a total of $B$ external gauge field and ghost lines becomes
\beq \label{71}
\om = 4 - B
\eeq 
which shows that only a finite number of combinations of external lines will yield divergent integrals. As a result the quantum gauge field theory of volume-preserving diffeomorphisms of ${\bf M}^{\sl 4}$ is renormalizable by power counting.

Let us finally consider the classification of asymptotic one-particle states assuming they are not confined which will be further analyzed in the next section.

To label the physical state-vectors we construct a basis of the one-particle Hilbert space of the gauge field theory of volume-preserving diffeomorphisms of ${\bf M}^{\sl 4}$ given by simultaneous eigenvectors of observables commuting amongst themselves as well as with the Hamiltonian of the theory. In other words we look for a complete system of conserved, commuting observables.

The specific difference of the present theory to a Yang-Mills theory arises from the structure of the gauge group - all observables not related to the gauge group remain the same and comprise the energy, the momentum and angular momentum three-vectors and other conserved internal degrees of freedom \cite{stw1}.

As the quantum gauge field theory of volume-preserving diffeomorphisms of ${\bf M}^{\sl 4}$ and minimally coupled "matter" field Lagrangians are translation and rotation invariant in inner space we have the additional conserved observables - the gravitational energy-momentum operator ${\bf K\/}_\al$ and the gravitational angular momentum tensor. As ${\bf K\/}_\al$ commutes with the already identified set of observables including the Hamiltonian (consistent with the Coleman-Mandula theorem) its eigenvalues $K_\al$ become additional quantum numbers labelling physical states. In addition, as all "matter" fields transform as scalars and the gauge and ghost fields as vectors under inner Lorentz transformations inner spin becomes yet another quantum number.

As a result we can find a basis of the one-particle Hilbert space 
\beq \label{72}
\mid k_\mu, \si; K_\al, \Si; \,\mbox{all other quantum numbers}\rangle
\eeq
labeled by the momentum four-vector $k_\mu$, the spin $\si$, the gravitational energy-momentum vector $K_\al$ and the inner spin $\Si$ which is $0$ for "matter" and $1$ for the gauge and ghost fields of the gauge field theory of volume-preserving diffeomorphisms of ${\bf M}^{\sl 4}$. The relation of these state vectors to asymptotic states describing observable particles with gravitational equal to inertial energy-momentum and the definition of the $S$-matrix is developed in detail in \cite{chw3}.

\section{Effective Action, Renormalization at One Loop and Asymptotic Freedom}
In this section starting from the formal perturbative expansion derived in the last section we calculate the renormalized effective action at one loop. The crucial point is to note that space-time and inner space integrals in the calculation of loop graphs completely decouple which allows us to first regularize the potentially divergent inner space integrals appropriately. Note that any consistent definition must respect the inner scale invariance of the classical action at the quantum level as this linearly realized symmetry is a symmetry of the quantum effective action as well \cite{stw2}. This allows us second to deal in the usual way with the ultraviolet divergencies related to the short distance behaviour in space-time and demonstrate the renormalizability of the gauge field theory of volume-preserving diffeomorphisms of ${\bf M}^{\sl 4}$ at one loop.

Technically we derive a formal expression for the one-loop effective action of the quantum gauge field theory of volume-preserving diffeomorphisms of ${\bf M}^{\sl 4}$ working in a covariant Minkowski background field gauge. We then define the inner momentum integrals using $\La$ as a cut-off and demonstrate the locality of the one-loop effective action in inner space. To prove the renormalizability at one loop we calculate the divergent contributions to the functional determinant of a general fluctuation operator with differential operator-valued coefficients in four space-time dimensions. Finally we determine the one-loop counterterms, renormalize the one-loop effective action and calculate the $\beta$-function of both the pure quantum gauge field theory of volume-preserving diffeomorphisms of ${\bf M}^{\sl 4}$ and the same theory minimally coupled to the Standard Model fields.

\subsection{Formal Expression}
To derive a formal expression for the one-loop effective action we work in a covariant Minkowski background field gauge choosing
\beq \label{73}
f^\ga [A,B] \equiv {_B\!{\cal D}}_\mu^\ga\,_\de A^{\mu \de} \eeq
where
\beq \label{74}
{_B\!{\cal D}}_\mu^\ga\,_\de
\equiv  \pa_\mu \,\eta^\ga\,_\de + B_\mu\,^\be\cdot \nabla_\be \,\eta^\ga\,_\de - \nabla_\de B_\mu\,^\ga \eeq
is the covariant derivative in the presence of a background field $B$ we will further specify below.

To get the one-loop expression for the generating functional Eqn.(\ref{46}) we have to expand the exponent around its stationary point up to second order in the fluctuations. Starting with
\beqq \label{75}
S_{MOD} [A, \om^*, \om; B] &=& -\frac{1}{4\, g^2\, \La^2} \,\int \,
F_{\mu\nu}\,^\al \cdot F^{\mu\nu}\,_\al \nonumber \\
&-& \frac{1}{2 \xi\, g^2\, \La^2} \,\int \,
{_B\!{\cal D}}^\mu_{\ga \al} A_\mu\,^\al \cdot {_B\!{\cal D}}_\nu^\ga\,_\be A^{\nu \be} \\
&+& \frac{1}{\La^2} \, \int \, \om^*_\ga \cdot {\cal F}^\ga\,_\de \left[A,B \right] \om^\de, \nonumber
\eeqq
where we have explicitly introduced a dimensionless gauge coupling $g^2$ and where
\beqq \label{76}
{\cal F}^\ga\,_\de \left[A, B \right] &=&
\frac{\de}{\de\, {\cal E}^\de} \,
{_B\!{\cal D}}_\mu^\ga\,\!_\al A_{_{\cal E}}^{\mu \al}\,_{\mid_{_{{\cal E}=0}}}
\nonumber \\
&=& {_B\!{\cal D}}_\mu^\ga\,\!_\al \, {\cal D}^{\mu \al}\,\!_\de \eeqq
is easily shown to be an endomorphism of ${\overline{\bf diff}}\,{\bf M}^{\sl 4}$ as required, we get the field equations in the presence of $J$ and $B$
\beqq \label{77}
& & {\cal D}_\mu^\be\,_\al F^{\mu\nu \al} 
+ \frac{1}{\xi} \, {_B\!{\cal D}}^{\nu \be}\,_\ga \, {_B\!{\cal D}}_\mu^\ga\,_\al 
A^{\mu \al} + g^2\, J^{\nu \be} = 0 \nonumber \\
& & {_B\!{\cal D}}_\mu^\ga\,_\al \, {\cal D}^{\mu \al}\,_\de \, \om^\de = 0 \\
& & {_B\!{\cal D}}^\mu_\de\,^\al \, {\cal D}_{\mu \al}\,^\ga \, \om^*_\ga = 0.
\nonumber \eeqq
They determine the stationary points $A_\mu\,^\al = A_\mu\,^\al [J, B]$, $\om^\de = 0$ and $\om^*_\ga = 0$ around which we expand. Setting the background field equal to the stationary point
\beq \label{78}
B_\mu\,^\al [J] =^{\!\!\!\!{!}}\: A_\mu\,^\al [J, B]
\eeq
determines $B$ as a functional of $J$ at least perturbatively.

Next we calculate the second variation of $S_{MOD}$ at the stationary points
\beqq \label{79}
\de^2 S_{MOD} &=& - \frac{1}{\La^2} \, \int \, \de A_\mu\,^\al \cdot
{\cal D}_{A,\xi}^{\mu\nu}\,_{\al\be}\, \de A_\nu\,^\be \nonumber \\
&-& \frac{2}{\La^2} \, \int \, \de \om^*_\ga \cdot {\cal D}_\om^\ga\,_\de\, \de \om^\de,
\eeqq
where we have absorbed the factors of $g$ in $\de A_\mu\,^\al$ and calculated the gauge and ghost field fluctuation operators to be
\beqq \label{80}
{\cal D}_{A,\xi}^{\mu\nu}\,_{\al\be} &\equiv& -\, \eta ^{\mu\nu} \cdot {\cal D}^\rho_\al\,^\ga \, {\cal D}_{\rho \ga\be} \, + \left(1 - \frac{1}{\xi} \right)\,
{\cal D}^\mu_\al\,^\ga \, {\cal D}^\nu_{\ga\be} \nonumber \\
& & -\,2\, F^{\mu\nu}\,_\ga \nabla^\ga \cdot \eta_{\al\be} 
+ 2\, \nabla_\be \, F^{\mu\nu}\,_\al \\
{\cal D}_\om^\ga\,_\de &\equiv& -\, {\cal D}^{\rho \ga\al} \, {\cal D}_{\rho \al\de}.
\nonumber \eeqq
They are endomorphisms of ${\overline{\bf diff}}\,{\bf M}^{\sl 4}$, i.e. if $f^\al$ fulfills $\nabla_\al f^\al =0$ so will ${\cal D}_{A,\xi}^{\mu\nu}\,_{\al\be} f^\be$ and ${\cal D}_\om^\ga\,_\de f^\de$, as is easily verified. Note that we had to commute ${\cal D}^\nu_{\ga\be}$ with ${\cal D}^\mu_\al\,^\ga$ to get the expression above for ${\cal D}_{A,\xi}^{\mu\nu}\,_{\al\be}$.

Taking all together we finally get
\beqq \label{81}
{\cal Z}_{\sl{1}-loop}\left[J \right]
&=& \int\Pi_{\!\!\!\!\!\!_{_{_{x,X;\mu,\al}}}}
\!\!\!\!\!\!\!\!d\,\de A_\mu\,^\al \; \Pi_{\!\!\!\!\!_{_{_{\mu}}}} \,\,
\de (\nabla_\al \de A_\mu\,^\al) \nonumber \\
& & \cdot \int\,\Pi_{\!\!\!\!\!\!_{_{_{x,X;\ga}}}}\!\!d\,\de\om^*_\ga \;
\de (\nabla^\ga \de\om^*_\ga)
\cdot \int\,\Pi_{\!\!\!\!\!\!_{_{_{x,X;\de}}}}\!\!d\,\de\om^\de \;
\de (\nabla_\de\, \de\om^\de) \nonumber \\
& & \cdot \exp\,i\,\left\{S_{MOD} [A, 0, 0; A] + \frac{1}{\La^2}\, \int J\cdot A \right\} \nonumber \\
& & \cdot \exp \Bigg\{ - \frac{i}{2\, \La^2} \, \int \, \de A_\mu\,^\al \cdot {\cal D}_{A,\xi}^{\mu\nu}\,_{\al\be}\, \de A_\nu\,^\be \\
& & \quad\quad\quad - \, \frac{i}{\La^2}\, \int \, \de \om^*_\ga \cdot {\cal D}_\om^\ga\,_\de\, \de \om^\de + \ep \mbox{-terms} \Bigg\} \nonumber \\
&=& \exp\,i\,\left\{S_{MOD} [A, 0, 0; A] + \frac{1}{\La^2}\, \int \, J\cdot A \right\} \nonumber \\
& & \cdot \Det^{-1/2}\, {\cal D}_{A,\xi} \cdot \Det\, {\cal D}_\om. \nonumber
\eeqq
As the fluctuation operators are endomorphisms of ${\overline{\bf diff}}\,{\bf M}^{\sl 4}$ the integrals in Eqn.(\ref{81}) are Gaussian and can be performed resulting in the usual determinants. Indeed, endowed with the scalar product Eqn.(\ref{45}), ${\overline{\bf diff}}\,{\bf M}^{\sl 4}$ becomes a Hilbert space with a complete orthonormal set of eigenvectors for each of the selfadjoint fluctuation operators above. These bases of the Hilbert space take the $\de$-functions automatically into account and the integration over each eigenvector direction becomes Gaussian.

Defining next the generating functional for connected Green functions
\beq \label{82}
{\cal W}\left[J \right] \equiv -\, i\, \Ln \, {\cal Z}\left[J \right]
\eeq
and the quantum effective action as its Legendre transform 
\beq \label{83}
\Ga\left[A \right] \equiv -\int J\cdot A + {\cal W}
\eeq
in the usual way we find
\beq \label{84}
\Ga_{\sl{1}-loop} \left[A \right]
= S_{MOD} [A, 0, 0; A] + \frac{i}{2} \Tr\Ln\, {\cal D}_{A,\xi} 
- \, i\, \Tr\Ln\, {\cal D}_\om
\eeq
which is the formal expression for the one-loop effective action we were looking for.

From now on we work with the specific choice $\xi = 1$ and drop the subscript $\xi$ to keep the calculations below as simple as possible.

\subsection{Finiteness and Locality of Inner Space Integrals}
To get a well-defined quantum theory at the one-loop level we have to show that the functional traces in Eqn.(\ref{84}) above evaluated over inner space can be appropriately defined, an issue which does not arise in Yang-Mills theories of compact Lie groups due to the finite volume of the underlying gauge groups.

To define $\Tr\Ln\, {\cal D}_A$ and $\Tr\Ln\, {\cal D}_\om$ and to demonstrate their locality in inner space we note that both operators are of the form
\beq \label{85} 
{\cal D} = -{\pa\rvec}^2 + {\cal M}_{\al\be}\,{\nabla\rvec}^\al{\nabla\rvec}^\be
+ {\cal N}_{\al}\,{\nabla\rvec}^\al + {\cal C},
\eeq
where ${\cal M}_{\al\be\mid \de}^{\,\ga}, {\cal N}_{\al\mid \de}^{\,\ga}, {\cal C}^\ga\,_\de$ are both matrices in inner space and matrix-valued differential operators in Minkowski space. This form is quite general and can account for covariant Minkowski background-field gauges such as in Eqn.(\ref{73}) as well, however, for $\xi\neq 1$ the operator would take an even more general form.

Properly normalizing and expanding the logarithm we obtain
\beqq \label{86}
\Tr\Ln \frac{{\cal D}}{{\cal D}_0}
&=& \Tr\Ln\, {\cal D} - \Tr\Ln\, {\cal D}_0 \nonumber \\
&=& \Tr\Ln\left({\bf 1} - \frac{1}{{\pa\rvec}^2}\,\left({\cal M}_{\al\be}\,{\nabla\rvec}^\al{\nabla\rvec}^\be
+ {\cal N}_{\al}\,{\nabla\rvec}^\al + {\cal C} \right) \right) \\
&=& \sum_n \frac{(-)^n}{n}\, \Tr\left[\left(- \frac{1}{{\pa\rvec}^2}\right)
\left({\cal M}_{\al\be}\,{\nabla\rvec}^\al{\nabla\rvec}^\be
+ {\cal N}_{\al}\,{\nabla\rvec}^\al + {\cal C}\right)\right]^n
\nonumber \\
&=& \sum_n \frac{(-)^n}{n}\,\Ga^{(n)}, \nonumber
\eeqq
where ${\cal D}_0$ is the operator for vanishing fields. Here we have defined the one-loop contribution with $n$ vertex insertions
\beqq \label{87}
\Ga^{(n)} &\equiv& \Tr_{\!\!\!\!\!\!\!\!\!_{_{_{x,X}}}}
\left[\left(- \frac{1}{{\pa\rvec}^2}\right)
\left({\cal M}_{\al\be}\,{\nabla\rvec}^\al{\nabla\rvec}^\be
+ {\cal N}_{\al}\,{\nabla\rvec}^\al + {\cal C}\right)\right]^n \nonumber \\
&=& \int\! d^{\sl 4} X_1\dots\dots\int\! d^{\sl 4} X_n
\int\! \frac{d^{\sl 4} P_1}{(2{\pi})^4}\dots\dots
\int\! \frac{d^{\sl 4} P_n}{(2{\pi})^4} \nonumber \\
& & \!\!\!\!\!\!\!\!\!\!\!\!\!\Tr_{\!\!\!\!\!\!\!_{_{_{x}}}}\: \Bigg\{ \langle X_1 \!\mid \left(- \frac{1}{{\pa\rvec}^2}\right)
\left({\cal M}_{\al_1 \be_1}\, {\nabla\rvec}^{\al_1}{\nabla\rvec}^{\be_1} + {\cal N}_{\al_1}\, {\nabla\rvec}^{\al_1} + {\cal C}\right) \mid\! P_1\rangle\,\cdot \nonumber \\
& & \quad\quad\quad\quad\quad\quad\quad\quad\quad\quad \vdots \nonumber \\
& & \cdot\langle X_n\!\mid \left(- \frac{1}{{\pa\rvec}^2}\right)
\left({\cal M}_{\al_n \be_n}\, {\nabla\rvec}^{\al_n}{\nabla\rvec}^{\be_n} + {\cal N}_{\al_n}\, {\nabla\rvec}^{\al_n} + {\cal C}\right) \mid\! P_n\rangle \Bigg\} \nonumber \\
& & \!\!\!\!\!\!\!\!\!\!\!\cdot \langle P_1\!\mid\! X_2\rangle\,\cdot\dots\cdot\,\langle P_n\!\mid\! X_1\rangle \\
&=& \int\! d^{\sl 4} X_1\dots\dots\int\! d^{\sl 4} X_n
\int\! \frac{d^{\sl 4} P_1}{(2{\pi})^4}\dots\dots
\int\! \frac{d^{\sl 4} P_n}{(2{\pi})^4} \nonumber \\
& & \!\!\!\!\!\!\!\!\!\!\!\!\!\Tr_{\!\!\!\!\!\!\!_{_{_{x}}}}\: 
\Bigg\{\left(- \frac{1}{{\pa\rvec}^2}\right)
\left({\cal M}_{\al_1 \be_1}\, {\nabla\rvec}^{\al_1}{\nabla\rvec}^{\be_1} + {\cal N}_{\al_1}\, {\nabla\rvec}^{\al_1} + {\cal C}\right)_{X_1} \,\cdot \nonumber \\
& & \quad\quad\quad\quad\quad\quad\quad\quad\quad\quad \vdots \nonumber \\
& & \cdot\left(- \frac{1}{{\pa\rvec}^2}\right)
\left({\cal M}_{\al_n \be_n}\, {\nabla\rvec}^{\al_n}{\nabla\rvec}^{\be_n} + {\cal N}_{\al_n}\, {\nabla\rvec}^{\al_n} + {\cal C}\right)_{X_n} \Bigg\} \nonumber \\
& & \!\!\!\!\!\!\!\!\!\!\!\cdot \exp \left(i P_1 (X_1 - X_2)
\,+ \dots +\, i P_n (X_n - X_1)\right) \nonumber 
\eeqq
which is manifestly invariant under local inner Poincar\'e transformations $X^\al\ar X'^\al = T^\al (x) + \La^\al\,_\be (x)\, X^\be$, $\La^\al\,_\be \in SO(1,3)$. Above we have inserted $n$ complete systems of both $X$- and $P$-vectors
\bed
{\bf 1} = \intX \mid\! X\rangle\langle X\!\mid, \quad\quad
{\bf 1} = \intP \mid\! P\rangle\langle P\!\mid
\eed
and used $\langle X\!\mid\! P\rangle = \exp(i\, P\cdot X)$ in Cartesian coordinates. Defining new variables
\beqq \label{88}
K_1 &\equiv& P_1 - P_n \nonumber \\
K_2 &\equiv& P_2 - P_1, \quad\quad\quad P_2 = K_2 + P_1 \nonumber \\
&\vdots& \quad\quad\quad\quad\quad\quad\quad\quad\:\: \vdots \\
K_{n-1} &\equiv& P_{n-1} - P_{n-2}, \quad P_{n-1}
= K_{n-1} + \dots + K_2 + P_1 \nonumber \\
K_n &\equiv& P_n - P_{n-1}, \quad\quad P_n
= K_n + \dots + K_2 + P_1 \nonumber
\eeqq
it becomes obvious that the definition of the $P_1$-integrals above over polynomials in $P_1$ requires care in order to avoid potential infinities related to the non-compactness of the gauge group.

We regularize such integrals generalizing our approach to define the classical action of the gauge field theory of volume-preserving diffeomorphisms of ${\bf M}^{\sl 4}$ in \cite{chw1} and get 
\beqq \label{89}
\Ga^{(n)}_{\La} &\equiv& \int\! d^{\sl 4}X_1\dots\dots\int\! d^{\sl 4}X_n
\intr \frac{d^{\sl 4}P_1}{(2{\pi})^4}
\int\! \frac{d^{\sl 4}K_2}{(2{\pi})^4} \dots\dots
\int\! \frac{d^{\sl 4}K_n}{(2{\pi})^4} \nonumber \\
& & \!\!\!\!\!\!\!\!\!\!\!\!\!\Tr_{\!\!\!\!\!\!\!_{_{_{x}}}}\:
\Bigg\{\left(- \frac{1}{{\pa\rvec}^2}\right)
\left({\cal M}_{\al_1 \be_1}\, i P_1^{\al_1} i P_1^{\be_1} +
{\cal N}_{\al_1}\, i P_1^{\al_1} + {\cal C}\right)_{X_1} \,\cdot \nonumber \\
& & \quad\quad\quad\quad\quad\quad\quad\quad\quad\quad \vdots \\
& & \!\!\!\!\!\!\!\!\!\!\!\!\!\!\!\!\cdot\left(- \frac{1}{{\pa\rvec}^2}\right)
\Big({\cal M}_{\al_n \be_n}\,
(i P_1^{\al_n} + i K_2^{\al_n} + \dots + i K_n^{\al_n})
(i P_1^{\be_n} + i K_2^{\be_n} + \dots + i K_n^{\be_n})
\nonumber \\
& & \quad\quad\quad +\:\: {\cal N}_{\al_n}\, (i P_1^{\al_n} + i K_2^{\al_n} + \dots + i K_n^{\al_n}) + {\cal C} \Big)_{X_n} \Bigg\} \nonumber \\
& & \!\!\!\!\!\!\!\!\!\!\!\cdot \exp \left( -i X_1 (K_2 + \dots + K_n)+ i X_2 K_2 \,+ \dots +\, i X_n K_n \right). \nonumber 
\eeqq

The regularization $\,\intr \frac{d^{\sl 4}P}{(2{\pi})^4}\dots$ consists in a) cutting off the Lorentz-invariant shells $- P^2 = M^2 < 0$ with negative mass squared consistent with the support condition Eqn.(\ref{2}) ensuring positive field energy, b) for a given non-negative mass squared cutting off the integrals over the Lorentz-invariant shells $- P^2 = M^2 \geq 0$ by requiring $P^{\sl 0} = \sqrt{M^2 + {\underline P}^2} \leq \frac{1}{2\,\La}$ for $P\in {\bf V^+}(P)$ and $P^{\sl 0} = -\sqrt{M^2 + {\underline P}^2} \geq -\frac{1}{2\,\La}$ for $P\in {\bf V^-}(P)$ and c) integrating over all non-negative $M^2$ which are bound by the cutoff for $P^{\sl 0}$ resulting in $M^2 \leq \frac{1}{4\, \La^2}$
\beqq \label{90}
& & \intr \frac{d^{\sl 4}P}{(2{\pi})^4}\dots
\equiv \int_0^{\frac{1}{4\, \La^2}}
\!\!\!\! dM^2\, \intP\, \de\!\left(M^2 + P^2 \right) \\
& & \quad \cdot\left(
\theta(P^{\sl 0}) \theta(-L^2 + 2\, L\cdot P) +
\theta(- P^{\sl 0}) \theta(-L^2 - 2\, L\cdot P)\right)\dots \nonumber
\eeqq
Note that this regularization respects the inner scale invariance. To write it in a Lorentz-covariant way we have used the fact that there is always a Lorentz frame with a time-like vector $L^\al$ which has $L^2 = -\La^{-2}$ as its invariant length so that $L^\al = (\La^{-1},\underline 0)$ in this frame.

Next, using $i K^\al_j \exp(i \sum_{l=2}^n X_l K_l) = {\nabla\rvec}^\al_j \exp(i \sum_{l=2}^n X_l K_l)$
and partially integrating we get
\beqq \label{91}
\Ga^{(n)}_\La &=& \int\! d^{\sl 4}X_1\dots\dots\int\! d^{\sl 4}X_n
\intr \frac{d^{\sl 4}P_1}{(2{\pi})^4}
\int\! \frac{d^{\sl 4}K_2}{(2{\pi})^4} \dots\dots
\int\! \frac{d^{\sl 4}K_n}{(2{\pi})^4} \nonumber \\
& & \!\!\!\!\!\!\!\!\!\!\!\!\!\Tr_{\!\!\!\!\!\!\!_{_{_{x}}}}\:
\Bigg\{\left(- \frac{1}{{\pa\rvec}^2}\right)
\left({\cal M}_{\al_1 \be_1}\, i P_1^{\al_1} i P_1^{\be_1} +
{\cal N}_{\al_1}\, i P_1^{\al_1} + {\cal C}\right)_{X_1} \,\cdot \nonumber \\
& & \quad\quad\quad\quad\quad\quad\quad\quad\quad\quad \vdots \\
& & \!\!\!\!\!\!\!\!\!\!\!\!\!\!\!\!\cdot\,\left(- \frac{1}{{\pa\rvec}^2}\right)
\Big({\cal M}_{\al_n \be_n}\,
(i P_1^{\al_n} - {\nabla\lvec}_2^{\al_n} - \dots - {\nabla\lvec}_n^{\al_n})
(i P_1^{\be_n} - {\nabla\lvec}_2^{\be_n} - \dots - {\nabla\lvec}_n^{\be_n}) \nonumber \\
& & \quad\quad\quad +\:\: {\cal N}_{\al_n}\, (i P_1^{\al_n} - {\nabla\lvec}_2^{\al_n} - \dots - {\nabla\lvec}_n^{\al_n}) + {\cal C} \Big)_{X_n} \Bigg\} \nonumber \\
& & \!\!\!\!\!\!\!\!\!\!\!\cdot \exp \left(i K_2 (X_2 - X_1) \,+ \dots +\, i K_n (X_n - X_1) \right). \nonumber 
\eeqq
Above, the differential operators act to the left and ordering obviously matters. Integrating over $K_i, X_j$ for $i,j=\sl{2,3}\dots n$ yields the final expression for $\Ga^{(n)}_\La $ in this subsection
\beqq \label{92}
\Ga^{(n)}_\La &=& \int\! d^{\sl 4}X_1
\intr \frac{d^{\sl 4}P_1}{(2{\pi})^4}
\nonumber \\
& & \!\!\!\!\!\!\!\!\!\!\!\!\!\Tr_{\!\!\!\!\!\!\!_{_{_{x}}}}\:
\Bigg\{\left(- \frac{1}{{\pa\rvec}^2}\right)
\left({\cal M}_{\al_1 \be_1}\, i P_1^{\al_1} i P_1^{\be_1} +
{\cal N}_{\al_1}\, i P_1^{\al_1} + {\cal C}\right)_{X_1} \,\cdot \nonumber \\
& & \quad\quad\quad\quad\quad\quad\quad\quad\quad\quad \vdots \\
& & \!\!\!\!\!\!\!\!\!\!\!\!\!\!\!\!\cdot\,\left(- \frac{1}{{\pa\rvec}^2}\right)
\Big({\cal M}_{\al_n \be_n}\,
(i P_1^{\al_n} - {\nabla\lvec}_2^{\al_n} - \dots - {\nabla\lvec}_n^{\al_n})
(i P_1^{\be_n} - {\nabla\lvec}_2^{\be_n} - \dots - {\nabla\lvec}_n^{\be_n}) \nonumber \\
& & \quad\quad\quad +\:\: {\cal N}_{\al_n}\, (i P_1^{\al_n} - {\nabla\lvec}_2^{\al_n} - \dots - {\nabla\lvec}_n^{\al_n}) + {\cal C} \Big)_{X_n=X_{n-1}=..=X_1} \Bigg\}. \nonumber
\eeqq
The expression above for $\Ga^{(n)}_\La $ is not only finite as an integral over inner space, but also local in $X_1$. Note that the regularized integrals over $P_1$ collapse into sums over products of metric tensors $\eta$ in inner space and factors of $\La$ to some power ensuring the correct dimension in inner space. These sums correspond to the sums over structure constants in the Yang-Mills case.

As in the case of the classical Lagrangian the contributions $\Ga^{(n)}$ to the one-loop effective action for $\rho\La$ are related to the ones for a given $\La$ by
\beq \label{93}
\Ga^{(n)}_{\rho \La} (\rho X,\rho A_\nu\,^\al (X),\dots) =
\Ga^{(n)}_{\La} (X,A_\nu\,^\al (X),\dots)
\eeq
respecting the scale invariance of the classical theory as they have to because this invariance is linearly realized and hence an invariance of the quantum effective action as well \cite{stw2}.

At one loop the dependence of the theory on $\La$ is again controlled by its scale invariance. In other words up to one loop theories for different $\La$ are equivalent up to inner rescalings. This symmetry is not distroyed by the renormalization required for the divergent space-time integrals with which we deal in the next subsection for the simple fact that both types of integrals and how we properly define them completely decouple.

\subsection{Divergence Structure of Space-time Integrals}
We turn to calculate the divergent contributions to the functional determinant of a general fluctuation operator with differential operator-valued coefficients in four space-time dimensions in preparation of the one-loop renormalization in the next subsection. 

To analyze the space-time divergencies occurring in $\Tr_\La \Ln \, {\cal D}_A$ and $\Tr_\La \Ln\, {\cal D}_\om$ we note that both operators are of the form
\beq \label{94} 
{\cal D} = -{\pa\rvec}^2 + {\cal B}_\rho\,{\pa\rvec}^\rho + {\cal C},
\eeq
where ${\cal B}_\rho, {\cal C}$ are both matrices in Minkowski space and matrix-valued differential operators in inner space. Again, this form is general enough to cope with covariant Minkowski background-field gauges such as in Eqn.(\ref{73}), however, the case $\xi\neq 1$ is not included.

Properly normalizing and expanding the logarithm we obtain
\beqq \label{95}
\Tr_\La \Ln \frac{{\cal D}}{{\cal D}_0}
&=& \Tr_\La \Ln\, {\cal D} - \Tr_\La \Ln\, {\cal D}_0 \nonumber \\
&=& \Tr_\La \Ln\left({\bf 1} - \frac{1}{{\pa\rvec}^2}\,\left({\cal B}_\rho\,{\pa\rvec}^\rho + {\cal C}\right) \right) \\
&=& \sum_n \frac{(-)^n}{n}\, \Tr_\La \left[\left(- \frac{1}{{\pa\rvec}^2}\right)
\left({\cal B}_\rho\,{\pa\rvec}^\rho + {\cal C}\right)\right]^n
\nonumber \\
&=& \sum_n \frac{(-)^n}{n}\,\Ga^{(n)}_\La, \nonumber
\eeqq
where ${\cal D}_0$ is the operator for vanishing fields. Here we have defined
\beqq \label{96}
\Ga^{(n)}_\La &\equiv& {\Tr_{\!\!\!\!\!\!\!\!\!_{_{_{x,X}}}}}_\La 
\left[\left(- \frac{1}{{\pa\rvec}^2}\right)
\left({\cal B}_\rho\,{\pa\rvec}^\rho + {\cal C}\right)\right]^n \nonumber \\
&=& \int\! d^4 x_1\dots\dots\int\! d^4 x_n
\int\! \frac{d^4 p_1}{(2{\pi})^4}\dots\dots
\int\! \frac{d^4 p_n}{(2{\pi})^4} \nonumber \\
& & \!\!\!\!\!{\Tr_{\!\!\!\!\!\!\!_{_{_{X}}}}}_\La \: \Bigg\{ \langle x_1\!\mid \left(- \frac{1}{{\pa\rvec}^2}\right)
\left({\cal B}_{\rho_1}\,{\pa\rvec}_1^{\rho_1} + {\cal C}\right) \mid \!p_1\rangle\,\cdot \nonumber \\
& & \quad\quad\quad\quad\quad\quad\quad \vdots \nonumber \\
& & \quad\quad \cdot\langle x_n\!\mid \left(- \frac{1}{{\pa\rvec}^2}\right)
\left({\cal B}_{\rho_n}\,{\pa\rvec}_n^{\rho_n} + {\cal C}\right) \mid \!p_n\rangle \Bigg\} \nonumber \\
& & \!\!\!\cdot \langle p_1\!\mid\! x_2\rangle\,\cdot\dots\cdot\,\langle p_n\!\mid\! x_1\rangle \\
&=& \int\! d^4 x_1\dots\dots\int\! d^4 x_n
\int\! \frac{d^4 p_1}{(2{\pi})^4}\dots\dots
\int\! \frac{d^4 p_n}{(2{\pi})^4} \nonumber \\
& & \!\!\!\!\!{\Tr_{\!\!\!\!\!\!\!_{_{_{X}}}}}_\La \: \Bigg\{
\frac{1}{p_1^2} \,
\left(i\,{\cal B}_{\rho_1}\,p_1^{\rho_1} + {\cal C}\right)_{x_1} \cdot
\nonumber \\
& & \quad\quad\quad\quad\quad\quad\quad \vdots \nonumber \\
& & \quad\quad \cdot \frac{1}{p_n^2} \,
\left(i\,{\cal B}_{\rho_n}\,p_n^{\rho_n} + {\cal C}\right)_{x_n}\Bigg\} \nonumber \\
& & \!\!\!\cdot \exp \left(i p_1 (x_1 - x_2) \,+ \dots +\, 
i p_n (x_n - x_1)\right), \nonumber 
\eeqq
where we have inserted $n$ complete systems of both $x$- and $p$-vectors
\bed
{\bf 1} = \intx \mid \!x\rangle\langle x\!\mid, \quad\quad
{\bf 1} = \intp \mid \!p\rangle\langle p\!\mid
\eed
and where $\langle x\!\mid\! p\rangle = \exp(i\, p\cdot x)$. Note the occurrence of the propagators above which is in marked difference to the local inner space integrals analyzed in the last subsection.

A shift of variables
\beqq \label{97}
k_1 &\equiv& p_1 - p_n \nonumber \\
k_2 &\equiv& p_2 - p_1, \quad\quad\quad p_2 = k_2 + p_1 \nonumber \\
&\vdots& \quad\quad\quad\quad\quad\quad\quad\quad \vdots \\
k_{n-1} &\equiv& p_{n-1} - p_{n-2}, \quad p_{n-1}
= k_{n-1} + \dots + k_2 + p_1 \nonumber \\
k_n &\equiv& p_n - p_{n-1}, \quad\quad p_n
= k_n + \dots + k_2 + p_1 \nonumber
\eeqq
allows us to rewrite $\Ga^{(n)}_\La$ as
\beqq \label{98}
\Ga^{(n)}_\La &=& \int\! d^4 x_1\dots\dots\int\! d^4 x_n
\int\! \frac{d^4 p_1}{(2{\pi})^4}\!
\int\! \frac{d^4 k_2}{(2{\pi})^4} \dots\dots
\int\! \frac{d^4 k_n}{(2{\pi})^4} \nonumber \\
& & \!\!\!\!\!\!\!\!\!\!{\Tr_{\!\!\!\!\!\!\!_{_{_{X}}}}}_\La \: \Bigg\{
\frac{1}{p_1^2} \,
\left(i\,{\cal B}_{\rho_1}\,p_1^{\rho_1} + {\cal C}\right)_{x_1} \cdot
\nonumber \\
& & \quad\quad\quad\quad\quad\quad\quad\quad\quad\quad \vdots \\
& & \!\!\!\!\!\cdot \frac{1}{(p_1 + k_2 + \dots + k_n)^2} \,
\left(i\,{\cal B}_{\rho_n}\,\left(p_1^{\rho_n} + k_2^{\rho_n} + \dots + k_n^{\rho_n}\right) + {\cal C}\right)_{x_n}\Bigg\} \nonumber \\
& & \!\!\!\!\!\!\!\!\cdot \exp \left( -i x_1 (k_2 + \dots + k_n)+ i x_2 k_2 \,+ \dots +\, i x_n k_n \right). \nonumber 
\eeqq
Now it is easy to read off the degrees of divergence $\om_n$ for the $p_1$-integrals which are bound by $\om_n\leq 4-n$. Hence, only the $\Ga^{(n)}_\La$ for $n=1,2,3,4$ have a divergent contribution.

Using dimensional regularization to isolate the divergent contributions which are local in $x_1$ we find
\beqq \label{99}
\left(\Tr_\La \Ln \frac{{\cal D}}{{\cal D}_0}\right)^{div} &=&
\Ga^{(1)\, div}_\La - \frac{1}{2}\, \Ga^{(2)\, div}_\La 
+ \frac{1}{3}\, \Ga^{(3)\, div}_\La - \frac{1}{4}\, \Ga^{(4)\, div}_\La \nonumber \\
&=& i\,\frac{\Om_4}{\ep} \intx_1 \,
{\Tr_{\!\!\!\!\!\!\!_{_{_{X}}}}}_\La \: \Bigg\{
- \frac{1}{12}\, \pa^\mu\,{\cal B}_\mu \cdot \pa^\nu\,{\cal B}_\nu \nonumber \\
&-& \frac{1}{24}\, \pa^\nu\,{\cal B}_\mu \cdot \pa_\nu\,{\cal B}^\mu + \frac{1}{2}\, \pa^\mu\,{\cal B}_\mu \cdot {\cal C}
- \frac{1}{2}\, {\cal C}^2 \\
&+& \frac{1}{12}\, \pa^\mu\,{\cal B}_\mu \cdot {\cal B}^\nu \cdot {\cal B}_\nu 
- \frac{1}{12}\, {\cal B}_\mu \cdot \pa^\mu {\cal B}^\nu \cdot {\cal B}_\nu \nonumber \\
&-& \frac{1}{4}\, {\cal C} \cdot {\cal B}^\nu \cdot {\cal B}_\nu
- \frac{1}{48}\, {\cal B}^\mu \cdot {\cal B}_\mu \cdot {\cal B}^\nu \cdot {\cal B}_\nu \nonumber \\
&-& \frac{1}{96}\, {\cal B}^\mu \cdot {\cal B}^\nu \cdot {\cal B}_\mu \cdot {\cal B}_\nu
\Bigg\}. \nonumber 
\eeqq
Above, we have used the results from Appendix A in \cite{chwB} for the $\Ga^{(n)\, div}_\La$ for $n=1,2,3,4$ with $\ep = d -4$ and $\Om_4 = \frac{1}{8\pi^2}$.

For fluctuations operators of the form
\beq \label{100} 
{\cal D} = -\, {\cal D}_\mu \,{\cal D}^\mu + {\cal E},\quad\quad
{\cal D}_\mu \equiv \pa_\mu + {\cal A}_\mu, 
\eeq
where the gauge field ${\cal A}_\mu$ is a matrix-valued differential operator, we have
\beq \label{101} 
{\cal B}_\mu = -\,2\, {\cal A}_\mu,\quad\quad {\cal C} = -\,\pa_\mu {\cal A}^\mu - {\cal A}_\mu \cdot {\cal A}^\mu + {\cal E}
\eeq
and using the cyclicality property of the trace, which is easily demonstrated, Eqn.(\ref{99}) further simplifies
\beq \label{102} 
\left(\Tr_\La \Ln \frac{{\cal D}}{{\cal D}_0}\right)^{div}
= - i\,\frac{\Om_4}{\ep} \intx_1 \,
{\Tr_{\!\!\!\!\!\!\!_{_{_{X}}}}}_\La \: \Bigg\{
\frac{1}{12}\, {\cal F}_{\mu\nu} \cdot {\cal F}^{\mu\nu} 
+ \frac{1}{2}\, {\cal E}^2 \Bigg\}.
\eeq
Above we have introduced the field strength operator 
\beq \label{103} 
{\cal F}_{\mu\nu} \equiv \left[{\cal D}_\mu, {\cal D}_\nu \right]
\eeq
which belongs to the gauge field operator ${\cal A}_\mu$.

\subsection{One-Loop Renormalization}
With the formulae Eqns.(\ref{102}) and (\ref{103}) which hold true for general fields living on both space-time and inner space we are now in a position to analyze the one-loop renormalizability of the quantum gauge field theory of volume-preserving diffeomorphisms of ${\bf M}^{\sl 4}$ both in the absence and presence of "matter" fields. Note that after properly regularizing the inner space integrals we can safely interchange the order of taking the traces over inner space versus space-time variables if needed. In this section we perform the functional trace over space-time variables first.

To analyze renormalizability we have to evaluate the divergent contributions to the one-loop effective action $\Ga_{\La, \sl{1}-loop} \left[A \right]$ in Eqn.(\ref{84}). A short calculation shows that the fluctuation operators Eqns.(\ref{80}) take the form of Eqn.(\ref{100}) above with
\beqq \label{104} 
\left({\cal A}_\mu\right)^\al\,_\be &=&  A_\mu\,^\ga \nabla_\ga \,\eta^\al\,_\be
- \nabla_\be A_\mu\,^\al \nonumber \\
\left({\cal F}_{\mu\nu}\right)^\al\,_\be &=&  F_{\mu\nu}\,^\ga \nabla_\ga \,\eta^\al\,_\be - \nabla_\be F_{\mu\nu}\,^\al \\
{\cal D}_A^{\mu\nu}\,_{\al\be} &=& -\, \eta ^{\mu\nu} \cdot \left({\cal D}^\rho\right)_\al\,^\ga \, \left({\cal D}_\rho\right)_{\ga\be} \, 
- 2\, \left({\cal F}_{\mu\nu}\right)_{\al\be} \nonumber \\
{\cal D}_\om^\ga\,_\de &=& -\, \left({\cal D}^\mu\right)^{\ga\al} \, \left({\cal D}_\mu\right)_{\al\de}.
\nonumber \eeqq
Taking the trace over space-time Minkowski indices we get the divergent contributions to the gauge field determinant in $d = 4+\ep$ dimensions
\beqq \label{105} 
\left(\Tr_\La \Ln \frac{{\cal D}_A}{{\cal D}_0}\right)^{div}
&=& - i\,\frac{\Om_4}{\ep} \intx \,
{\Tr_{\!\!\!\!\!\!\!_{_{_{X}}}}}_\La \: \Bigg\{
\frac{1}{12}\,4\, {\cal F}_{\mu\nu} \cdot {\cal F}^{\mu\nu} 
+ \frac{1}{2}\,4\, {\cal F}_{\mu\nu} \cdot {\cal F}^{\nu\mu} \Bigg\}
\nonumber \\
&=& i\,\frac{\Om_4}{\ep} \,\frac{5}{3}\,4\, \intx 
{\Tr_{\!\!\!\!\!\!\!_{_{_{X}}}}}_\La \: F_{\mu\nu} \cdot F^{\mu\nu},
\eeqq
and to the ghost determinant
\beqq \label{106}
\left(\Tr_\La \Ln \frac{{\cal D}_\om}{{\cal D}_0}\right)^{div}
&=& - i\,\frac{\Om_4}{\ep} \intx \,
{\Tr_{\!\!\!\!\!\!\!_{_{_{X}}}}}_\La \:
\frac{1}{12} \, {\cal F}_{\mu\nu} \cdot {\cal F}^{\mu\nu} \nonumber \\
&=& - i\,\frac{\Om_4}{\ep} \,\frac{1}{12}\,4\, \intx 
{\Tr_{\!\!\!\!\!\!\!_{_{_{X}}}}}_\La \:  F_{\mu\nu} \cdot F^{\mu\nu}.
\eeqq
Note that as for other gauge field theories it is the second term in Eqn.(\ref{105}) which determines the sign of the gauge field contribution above - which will in turn determine the sign of the $\beta$-function of the quantum gauge field theory of volume-preserving diffeomorphisms of ${\bf M}^{\sl 4}$.

Taking all together we find
\beqq \label{107}
\Ga_{\La, \sl{1}-loop}^{div} \left[A \right]
&=&
\frac{i}{2} \left(\Tr_\La \Ln \frac{{\cal D}_A}{{\cal D}_0}\right)^{div}
- \, i\, \left(\Tr_\La \Ln \frac{{\cal D}_\om}{{\cal D}_0}\right)^{div} \nonumber \\
&=& - \,\frac{\Om_4}{\ep} \,\frac{11}{3}\, \intx {\Tr_{\!\!\!\!\!\!\!_{_{_{X}}}}}_\La \:  F_{\mu\nu} \cdot F^{\mu\nu} \\
&=& - \,\frac{\Om_4}{\ep}\, \frac{11}{3}\, \Om^\La_{\sl 1}\, 
\frac{1}{\La^2} \,\int F_{\mu\nu}\,^\al \cdot F^{\mu\nu}\,_\al,
\nonumber \eeqq
where $\Om^\La_{\sl 1} = \frac{1}{720\, (4\pi)^3}$ as calculated in \cite{chw1}. The one-loop divergence is proportional to the original action of the gauge field theory of volume-preserving diffeomorphisms of ${\bf M}^{\sl 4}$ and the theory is renormalizable at one loop. Note the formal similarity of the formula above with the analogous expression for Yang-Mills theories, especially the occurrence of the universal numerical factor $\frac{11}{3}$.

As usual the divergent contribution $\Ga_{\La, \sl{1}-loop}^{div} \left[A \right]$ can be absorbed in the original action of the gauge field theory of volume-preserving diffeomorphisms through a redefinition of the gauge coupling constant
\beq \label{108}
g_R = g\left(1 - \frac{g^2}{180\, (4 \pi)^5}\, \frac{11}{3}\, \frac{1}{\ep} + O(g^4)\right)
\eeq
where we have used $\Om_4=\frac{1}{8\pi^2}$. 

As a result the one-loop effective action after regularization of the inner space integrals and renormalization is a perfectly well defined expression.

The corresponding $\be$-function of the gauge field theory of volume-preserving diffeomorphisms at one loop becomes
\beq \label{109}
\be(g) = - \frac{g^3}{180\, (4 \pi)^5}\, \frac{11}{3} 
\eeq
and the theory is asymptotically free.

Note that $\La$ does not get renormalized as we would expect from the complete decoupling of inner and space-time integrals and their treatments.

\subsection{Inclusion of Standard Model "Matter" Fields}
As discussed in \cite{chw1} the fields of the gauge theory of volume-preserving diffeomorphisms of ${\bf M}^{\sl 4}$ interact with {\it all} fundamental fields appearing in a QFT such as the SM of elementary particle physics through minimal coupling. For clarity we call all these other fundamental scalar, spinor and (gauge) vector fields "matter" fields in the sequel. For a potential physical interpretation of the gauge field theory of volume-preserving diffeomorphisms of ${\bf M}^{\sl 4}$ it is thence crucial to extend the analysis of the asymptotic scaling behaviour above to include the impact of these other fields on the renormalized coupling and the $\be$-function.

To be specific let us do this analysis for the SM fields which we mini-mally couple to the gauge field theory of volume-preserving diffeomorphisms by (1) allowing all SM fields to live on ${\bf M\/}^{\sl 4}\times {\bf M\/}^{\sl 4}$ (with possibly the same restriction for the support of inner space Fourier-transformed "matter" fields to ${\bf V^+}(K)\cup {\bf V^-}(K))$ - adding the necessary additional inner degrees of freedom -  and by (2) replacing ordinary derivatives through covariant ones $\pa_\mu\ar D_\mu = \pa_\mu + A_\mu\,^\al\cdot \nabla_\al $ in all "matter" Lagrangians as usual. 

In Appendix A we have derived the additional divergent contributions ${\it\Delta}\Ga_{\La, \sl{1}-loop}^{div} \left[A \right]$ to the one-loop effective action contributing to the renormalization of the gauge field theory of volume-preserving diffeomorphisms of ${\bf M}^{\sl 4}$.

To apply this to the SM let us recall its field content. The SM is built by gauging $SU(3)\times SU(2)\times U(1)$ which leaves us with $8$ strongly, $3$ weakly and $1$ electromagnetically interacting gauge fields - $12$ in total. These fields interact with $3$ families of leptons and quarks, two of which are structural replications of the first family consisting of the $15$ chiral Dirac fields for $\nu_e, e_L, e_R, u^a_L, u^a_R, d^a_L, d^a_R$, where $a = 1,2,3$ indicates the strongly interacting color degrees of freedom. Finally there is a Higgs dublett adding two scalar degrees of freedom.

In total we have (see Appendix A for the derivation)
\beqq \label{110}
& & \Ga_{\La, \sl{1}-loop}^{div} \left[A \right] \ar 
\Ga_{\La, \sl{1}-loop}^{div} \left[A \right] +
\,12\, {\it\Delta}_G \! \Ga_{\La, \sl{1}-loop}^{div} \left[A \right]
\nonumber \\
& & \quad\quad + \,\,45\, {\it\Delta}_D \! \Ga_{\La, \sl{1}-loop}^{div} \left[A \right] + \,2\, {\it\Delta}_S \! \Ga_{\La, \sl{1}-loop}^{div} \left[A \right] \\
& & = - \,\frac{\Om_4}{\ep}\, \frac{1}{12}\, 
\Big(44 + 24 - 90 - 2 \Big) \Om^\La_{\sl 1}\, 
\frac{1}{\La^2}\, \int \,F_{\mu\nu}\,^\al \cdot F^{\mu\nu}\,_\al, \nonumber
\eeqq
where $24$ is the contribution of the SM gauge fields, $90$ of the leptons and quarks and $2$ of the Higgs respectively. This translates into the renormalized coupling
\beq \label{111}
g_R = g\left(1 + \frac{g^2}{180\, (4 \pi)^5}\, 2\, \frac{1}{\ep} + O(g^4)\right)
\eeq
and the $\be$-function
\beq \label{112}
\be(g) = +\, \frac{g^3}{180\, (4 \pi)^5}\, 2 
\eeq
of the gauge field theory of volume-preserving diffeomorphisms of ${\bf M}^{\sl 4}$ minimally coupled to the Standard Model fields.

The combined theory is not asymptotically free and we expect the inner space degrees of freedom and the gauge and "matter" fields associated with them to be observable and asymptotic free field states to exist which we have discussed in detail in \cite{chw3}. In this case it also makes sense to evaluate the classical limit of the gauge field theory of volume-preserving diffeomorphisms of ${\bf M}^{\sl 4}$ as we have done in \cite{chw2} deriving Newton's inverse square law of gravitation.

\section{BRST Symmetry and BRST Quantization}
In this section we introduce the nilpotent BRST transformations for the gauge field theory of volume-preserving diffeomorphisms of ${\bf M}^{\sl 4}$ and establish the BRST invariance of the gauge-fixed action. We define the physical states as equivalence classes of states in the kernel of the nilpotent BRST operator $Q$ modulo the image of $Q$. Finally we discuss the generalized BRST quantization of the gauge field theory of volume-preserving diffeomorphisms of ${\bf M}^{\sl 4}$.

Let us start with the modified action $S_{MOD}$ from Eqn.(\ref{47}) which may be written as
\beq \label{113}
S_{MOD} = S - \frac{1}{2\xi\, \La^2}\, \int\,f_\ga \cdot f^\ga
+ \frac{1}{\La^2}\, \int\, \om^*_\ga \cdot {\it\Delta}^\ga,
\eeq
where we have introduced the quantity
\beq \label{114}
{\it\Delta}^\ga \equiv {\cal F}^\ga \,_\de \, \om^\de.
\eeq 
Next we re-express
\beqq \label{115}
B[f] &=& \exp\left\{ - i\, \frac{1}{2\xi\, \La^2}\, \int\, f_\ga \cdot f^\ga \right\} \\
&\propto& \int\Pi_{\!\!\!\!\!\!_{_{_{x,X;\ga}}}}\!\!\!\!dh^\ga \;
\de (\nabla_\ga h^\ga)
\cdot \exp\left\{ i\, \frac{\xi}{2\, \La^2}\, \int\,h_\ga \cdot h^\ga 
+ i\, \frac{1}{\La^2}\, \int\,h_\ga \cdot f^\ga \right\}
\nonumber \eeqq
as a Gaussian integral and introduce the corresponding new modified action
\beq \label{116}
S_{NEW} = S + \frac{1}{\La^2}\, \int\, \om^*_\ga \cdot {\it\Delta}^\ga + \frac{1}{\La^2}\, \int\,h_\ga \cdot f^\ga + \frac{\xi}{2\, \La^2} \, \int\,h_\ga \cdot h^\ga.
\eeq
Green functions are now given as path integrals over the fields $A$, $\om^*$, $\om$, $h$, $\psi$ with weight $\exp\, i \, \{S_{NEW}+S_M\}$.

By construction the gauge-fixed modified action $S_{NEW}$ is not invariant under gauge transformations. However, it is invariant under BRST transformations parametrized by an infinitesimal constant $\theta$ anticommuting with ghost and fermionic fields. The BRST variations are given by
\beqq \label{117}
\de_\theta A_\mu\,^\al &=& \theta\left(\pa_\mu \om^\al
+ A_\mu\,^\be \nabla_\be \om^\al -\om^\be \nabla_\be A_\mu\,^\al \right) \nonumber \\
\de_\theta \om^*_\ga &=& - \theta\, h_\ga \nonumber \\
\de_\theta \om^\de &=& - \theta\, \om^\be \nabla_\be \om^\de \\
\de_\theta h_\ga &=& 0 \nonumber \\
\de_\theta \psi &=& - \theta\, \om^\be \nabla_\be \psi.
\nonumber \eeqq
The transformations Eqns.(\ref{117}) are nilpotent, i.e. if ${\cal F}$ is any functional of $A, \om^*, \om, h, \psi$ and we define $s{\cal F}$ by
\beq \label{118}
\de_\theta {\cal F} \equiv \theta s{\cal F}
\eeq
then  
\beq \label{119}
\de_\theta s{\cal F} = 0 \quad\mbox{or}\quad s(s{\cal F}) = 0.
\eeq 
The proof for the fields above is straightforward, but somewhat tedious. Here we just give a sketch of the verification of $s(s A_\mu\,^\al)=0$
\beqq \label{120}
\de_\theta sA_\mu\,^\al &=& \theta\, \Bigg\{
\pa_\mu \left( - \om^\be \nabla_\be \om^\al \right) \nonumber \\
&+& \left(\pa_\mu \om^\be
+ A_\mu\,^\ga \nabla_\ga \om^\be -\om^\ga \nabla_\ga A_\mu\,^\be \right) \nabla_\be \om^\al \nonumber \\
&-& A_\mu\,^\be \nabla_\be \left( \om^\ga \nabla_\ga \om^\al \right)
+ \left( \om^\ga \nabla_\ga \om^\be \right) \nabla_\be A_\mu\,^\al \\
&+& \om^\be \nabla_\be \left( \pa_\mu \om^\al 
+ A_\mu\,^\ga \nabla_\ga \om^\al - \om^\ga \nabla_\ga A_\mu\,^\al \right)\Bigg\} \nonumber \\
&=& 0 \nonumber \eeqq
using the chain-rule and the anticommutativity of $\theta$ with $\om$. As a result we have
\beqq \label{121}
s(s A_\mu\,^\al) &=& 0,\quad s(s \om^*_\ga) = 0, \quad s(s \om^\de) = 0 \nonumber \\
s(s h_\ga) &=& 0, \quad s(s\psi) = 0.
\eeqq
The extension to products of polynomials in these fields follows then easily.

To verify the BRST invariance of $S_{NEW}$ we note that the BRST transformation acts on functionals of matter and gauge fields alone as a gauge transformation with gauge parameter ${\cal E}_\al = \theta\, \om_\al$. Hence
\beq \label{122}
\de_\theta S = 0.
\eeq
Next with the use of Eqn.(\ref{36}) we determine the BRST transform of $f^\ga$
\beq \label{123}
\de_\theta f^\ga = \frac{\de f^\ga}{\de\, {\cal E}_\al} _{\mid_{_{{\cal E}=0}}} \!\!\!\! \theta \, \om_\al= \theta \, {\it\Delta}^\ga
\eeq
which yields
\beq \label{124}
\om^*_\ga \cdot {\it\Delta}^\ga + h_\ga \cdot f^\ga + \frac{\xi}{2} \, h_\ga \cdot h^\ga = -s\left(\om^*_\ga \cdot f^\ga + \frac{\xi}{2} \, \om^*_\ga \cdot h^\ga \right).
\eeq
Hence we can rewrite
\beq \label{125}
S_{NEW} = S + s{\it\Psi},
\eeq
where
\beq \label{126}
{\it\Psi} \equiv -\frac{1}{\La^2}\, \int \left(\om^*_\ga \cdot f^\ga + \frac{\xi}{2} \, \om^*_\ga \cdot h^\ga \right).
\eeq
Finally it follows from the nilpotency of the BRST transformation
\beq \label{127}
\de_\theta S_{NEW} = 0.
\eeq

As for Yang-Mills theories Eqn.(\ref{127}) shows that the physical content of the gauge field theory of volume-preserving diffeomorphisms of ${\bf M}^{\sl 4}$  is contained in the kernel of the BRST transformation modulo terms in its image.

Equivalent to this is the requirement that matrix elements between physical states $\mid\!\! a\rangle,\dots$ are independent of the choice of the gauge-fixing functional ${\it\Psi}$. This implies the existence of a nilpotent BRST operator $Q$ with $Q^2=0$. Physical states are then in the kernel of $Q$
\beq \label{128}
Q \!\mid\! a\rangle = 0,\quad \langle b \!\mid\! Q = 0.
\eeq
Independent physical states are defined as the equivalence classes of states in the kernel of $Q$ modulo the image of $Q$.

Finally let us note that as for Yang-Mills theories \cite{stw2} we can generalize the Faddeev-Popov-de Witt quantization procedure. In the general case one starts with an action given as the most general local functional of $A$, $\om^*$, $\om$, $h$, $\psi$ with ghost number zero which is invariant under the BRST transformations Eqns.(\ref{117}) and any other global symmetry of the theory as well as with dimension less or equal to four so as to assure renormalizability. Such actions are of the general form \cite{stw2}
\beq \label{129}
S_{NEW} [A,\om^*,\om,h,\psi]  = S [\phi] + s{\it\Psi} [A,\om^*,\om,h,\psi]
\eeq
with $s{\it\Psi}$ being a general functional respecting the restrictions above.

$S$-matrix elements of physical states annihilated by the appropriate BRST operator of the theory are then independent of ${\it\Psi}$. In addition, in the Minkowski-plus-axial gauge the ghosts decouple in the quantum gauge field theory of volume-preserving diffeomorphisms of ${\bf M}^{\sl 4}$, hence they decouple for any choice of ${\it\Psi}$ and the {\it theory is ghost-free}.

\section{Renormalizability to All Orders}
In this section we sketch a proof of the renormalizability of the gauge field theory of volume-preserving diffeomorphisms of ${\bf M}^{\sl 4}$ to all orders.

A general proof of the renormalizability of the gauge field theory of volume-preserving diffeomorphisms of ${\bf M}^{\sl 4}$, i.e. the existence of a finite, well-defined perturbative effective action, has to comprise the analysis of the divergence structure and the renormalizability of space-time integrals as for Yang-Mills theories and in addition the verification that inner space integrals can be properly regularized respecting the scale invariance of the classical theory which is a key condition as this is a linearly realized classical symmetry which extends necessarily to the quantum effective action.

Turning to the first point we note that we should be able to employ the full machinery developed for the inductive proof of renormalizability for Yang-Mills gauge theories as the general structure of the quantum gauge field theory of volume-preserving diffeomorphisms of ${\bf M}^{\sl 4}$ formally is close to that of quantum Yang-Mills theories. Hence we should be able to repeat all the steps in the renormalizability proof e.g. given in the Chapters 15 to 17 in \cite{stw2} or in \cite{jez}. The only change arises from the slightly different form of the BRST transformations for the gauge field theory of volume-preserving diffeomorphisms of ${\bf M}^{\sl 4}$ as compared to Yang-Mills gauge theories requiring the adaptation of the analysis given in Section 17.2 of \cite{stw2}.

Turning to the second point our approach at the one-loop level has been to
(1) define the inner one-loop integrals using $\La$ as a scale-invariant cut-off as in Eqn.(\ref{90})
\beq \label{130}
\int \! \frac{d^{\sl 4}P_1}{(2{\pi})^4} \times \mbox{integrand} \ar
\intr \frac{d^{\sl 4}P_1}{(2{\pi})^4} \times \mbox{integrand}
\eeq
and (2) on the basis of this definition to demonstrate the validity of the scaling law
\beq \label{131}
\Ga^{(1-loop)}_{\rho \La} (\rho X,\rho A_\nu\,^M(X),\dots) =
\Ga^{(1-loop)}_{\La} (X,A_\nu\,^M(X),\dots) 
\eeq
ensuring consistency and the uniqueness of the theory up to inner rescalings.

The same strategy should work for any number of loops. Again (1) we {\it regularize} inner $n$-loop integrals arising in the calculation of the effective action by the scale-invariant prescription Eqn.(\ref{90})
\beqq \label{132}
& & \int \! \frac{d^{\sl 4}P_1}{(2{\pi})^4} \cdot\dots\cdot \! \int \! \frac{d^{\sl 4}P_n}{(2{\pi})^4} \times \mbox{integrand} \\
& & \quad \ar \, \intr \frac{d^{\sl 4}P_1}{(2{\pi})^4} \cdot\dots\cdot \intr \frac{d^{\sl 4}P_n}{(2{\pi})^4} 
\times \mbox{integrand}
\nonumber \eeqq
and (2) on the basis of this regularization we should be able to demonstrate the validity of the scaling law
\beq \label{133}
\Ga^{(n-loop)}_{\rho \La} (\rho X,\rho A_\nu\,^M(X),\dots) =
\Ga^{(n-loop)}_{\La} (X,A_\nu\,^M(X),\dots) 
\eeq
noting that the inner scale invariance is a {\it linearly realized} symmetry of the gauge field theory of volume-preserving diffeomorphisms of ${\bf M}^{\sl 4}$  and hence a symmetry of the quantum effective action \cite{stw2}. This ensures the uniqueness of the theory up to inner rescalings at $n$ loops.

The locality of the theory in inner space for any number of loops follows from the non-propagation of inner degrees of freedom which can be most easily read off the propagators in Eqns.(\ref{59}).

This completes the sketch of a general proof of the renormalizability and the essential uniqueness of the quantum gauge field theory of volume-preserving diffeomorphisms of ${\bf M}^{\sl 4}$.

\section{Conclusions}
In this paper we have quantized the classical gauge theory of volume-preserving diffeomorphisms of ${\bf M}^{\sl 4}$ in the path integral formalism starting with a Hamiltonian formulation of the theory with unconstrained, though neither Lorentz- nor gauge-covariantly looking canonical field variables and a manifestly positive Hamiltonian. As the canonical field variables obey the usual Poisson brackets the physical Hilbert space of states has positive norm and is ghost-free.

Over various steps we then have brought the relevant path integral measure and weight into a Lorentz- and gauge-covariant form allowing us to express correlation functions first in the Minkowski-plus-axial gauge and - applying the De Witt-Faddeev-Popov approach - in any meaningful gauge. On that basis we have developed the Feynman rules of the theory and demonstrated that the gauge theory of volume-preserving diffeomorphisms of ${\bf M}^{\sl 4}$ is renormalizable by power-counting. Finally we have discussed the new quantum numbers appearing in the theory which label state vectors. 

Next we have calculated and renormalized the divergent parts of the quantum effective action at one loop in a background field approach. Here we had to deal not only with the usual short distance divergencies of space-time integrals in a perturbation expansion \cite{stw1,stw2}, but - due to the non-compactness of the gauge group - also with additional divergent integrals over inner space. We have regularized these based on the requirement of respecting the relevant inner symmetries (inner Lorentz and most importantly inner scale invariance) - generalizing thereby the finite sums over structure constants appearing in the perturbation series for the Yang-Mills case to the present one. The result at one loop is a negative $\beta$-function and hence an asymptotically free theory without the presence of other fields and a positive $\beta$-function and a theory without asymptotic freedom after minimally coupling the Standard Model (SM) fields to the gauge fields.

Finally we have developed the BRST apparatus as preparation for the renormalizability proof to all orders and given a sketch of this proof which in itself is one of the open points to be further adressed. Yet we have demonstrated that the gauge theory of volume-preserving diffeomorphisms of ${\bf M}^{\sl 4}$ is a quantum field theory fulfilling the key requirements towards a physical theory: namely to have a positive Hamiltonian, a ghost free Hilbert space of states with positive norm and a unitary S-matrix. Taking this into account together with the demonstration that the theory at the classical level yields a relativistic description of gravitation we propose the gauge theory of volume-preserving diffeomorphisms of ${\bf M}^{\sl 4}$ as a viable candidate for a renormalizable quantum theory of gravity.

On top, none of the well-known fundamental difficulties such as the disappearance of the notion of a particle or the non-existence of non-trivial correlators arising in the attempts to quantize General Relativity or any other geometric theory of gravity \cite {car,clk} plagues the current approach as all the notions developed in the context of a relativistic QFT can immediately be generalized to our context.

Also let us point to interesting not yet analysed questions such as to the structure of the vacuum of the present theory - noting that unlike in the Yang-Mills case, where $F=0$ implies that $A$ is pure gauge, in our case $F=0$ also results from any $A=constant$ - or to the perturbative calculation of correlation functions and scattering cross-sections which in the gravitational scattering of matter should result in the non-relativistic limit in well-known Rutherford-type formulae \cite{cli} allowing for further consistency checks.

Finally what makes the gauge theory of volume-preserving diffeomorphisms of ${\bf M}^{\sl 4}$ an attractive candidate for a consistent classical and quantum theory of gravity in the first place is its structural analogy with the existing gauge field theories of the electromagnetic, weak and strong interactions \cite{lor,stp,tpc}. If it was the "right" theory we would finally have a unified view of Nature and a consistent framework to describe all fundamental interactions at all accessible scales and without any logical or mathematical rift between the worlds of classical and quantum physics.

\appendix

\section{"Matter" Contributions to Divergent Part of One-Loop Effective Action of the Gauge Field Theory of Volume-Preserving Diffeomorphisms of ${\bf M}^{\sl 4}$}
In this Appendix we calculate the divergent vacuum contribution of a gauge vector field, a Dirac spinor and a complex scalar doublet to the one-loop effective action of the gauge field theory of volume-preserving diffeomorphisms of ${\bf M}^{\sl 4}$.

\subsection{Gauge field contribution ${\it\Delta}_G \! \Ga_{\La, \sl{1}-loop}^{div} \left[A \right]$}
The vacuum amplitude of a Yang-Mills gauge field $B_\mu\,^a$ with gauge algebra indices $a, b,..=1,..,\dim A$ minimally coupled to the gauge field theory of volume-preserving diffeomorphisms of ${\bf M}^{\sl 4}$, where $\dim A$ is the dimension of the gauge algebra, is given by
\beqq \label{134}
{\cal Z}_G [A] &\equiv&
\int\Pi_{\!\!\!\!\!\!_{_{_{x,X;\mu,a}}}}
\!\!\!\!\!\!\!\!dB_\mu\,^a \;
\int\Pi_{\!\!\!\!\!\!_{_{_{x,X;b}}}}\!\!d\om^*_b \;
\int\Pi_{\!\!\!\!\!\!_{_{_{x,X;c}}}}\!\!d\om^c
\nonumber \\
& & \cdot \exp\,i\,\Big\{S_{MOD} + \ep \mbox{-terms} \Big\}, \eeqq
where $D_\mu^a\,\!_b [B] = \pa_\mu \de^a\,\!_b + C^a\,_{cb} B_\mu\,^c$ is the covariant derivative in the presence of a gauge field $B$, $C^a\,_{cb}$ the structure constants of the gauge algebra and $\om^*_b$, $\om^c$ the ghost fields corresponding to the gauge-fixed action $S_{MOD}$
\beqq \label{135}
S_{MOD}&\equiv& S_{YM} + S_{GF} + S_{GH} \nonumber \\
S_{YM}&\equiv& -\frac{1}{4} \,\int \, {\overline G}_{\mu\nu}\,^a \cdot {\overline G}^{\mu\nu}\,_a \\
S_{GF}&\equiv& - \frac{1}{2\xi} \,\int \,
{\overline D}^\mu_{ab} [C] B_\mu\,^b \cdot {\overline D}_\nu^a\,\!_c [C] B^{\nu c} \nonumber \\
S_{GH}&\equiv& \int \, \om^*_b \cdot {\cal F}^b\,_c \left[B, C \right] \om^c. \nonumber
\eeqq
$C_\mu\,^a$ appearing in the gauge-fixing and ghost terms is a background gauge field. Above we have minimally coupled the Yang-Mills field to the gauge field theory of volume-preserving diffeomorphisms of ${\bf M}^{\sl 4}$ replacing ordinary through covariant derivatives $\pa_\mu \ar D_\mu = \pa_\mu + A_\mu\,^\al\cdot \nabla_\al$ yielding
\beq \label{136}
D_\mu^a\,\!_b [B] \ar {\overline D}_\mu^a\,\!_b [B] = D_\mu \de^a\,\!_b + C^a\,_{cb} B_\mu\,^c,
\eeq
and introduced the field strength and the ghost fluctuation operator
\beqq \label{137}
{\overline G}_{\mu\nu}\,^a &=& D_\mu B_\nu\,^a - D_\nu B_\mu\,^a +
C^a\,_{bc} \, B_\mu\,^b \, B_\nu\,^c, \nonumber \\
{\cal F}^b\,_c \left[B, C \right] &=&
{\overline D}_\mu^b\,\!_a [C] \, {\overline D}^{\mu a}\,\!_c [B].
\eeqq
The bars over derivatives etc. indicate minimal coupling to the gauge field theory of volume-preserving diffeomorphisms of ${\bf M}^{\sl 4}$.

Expanding $S_{MOD}$ around its stationary points $B_\mu\,^a = C_\mu\,^a  = \om^*_b = \om^c = 0$ in the absence of source terms and performing the Gaussian integral gives
\beqq \label{138}
{\cal Z}_{G, \sl{1}-loop} [A]
&=& \int\Pi_{\!\!\!\!\!\!_{_{_{x,X;\mu,a}}}}
\!\!\!\!\!\!\!\!d\de B_\mu\,^a \;
\int\Pi_{\!\!\!\!\!\!_{_{_{x,X;b}}}}\!\!d\de \om^*_b \;
\int\Pi_{\!\!\!\!\!\!_{_{_{x,X;c}}}}\!\!d\de \om^c \nonumber \\
& & \cdot \exp \Bigg\{ - \frac{i}{2} \, \int \, \de B_\mu\,^a \cdot
{\cal D}_{B,\xi}^{\mu\nu}\,_{ab}\, \de B_\nu\,^b \\
& & \quad\quad - \: \, \int \, \de \om^*_b \cdot {\cal D}_\om^b\,_c\, \de \om^c \Bigg\} \nonumber \\
&=& \Det^{-1/2}\, {\cal D}_{B,\xi} \cdot \Det\, {\cal D}_\om, \nonumber
\eeqq
where
\beqq \label{139}
{\cal D}_{B,\xi}^{\mu\nu}\,_{ab} &\equiv& - \left(\eta ^{\mu\nu} \cdot D^\rho \, D_\rho \, + \left(1 - \frac{1}{\xi} \right)\,
D^\mu \, D^\nu - F^{\mu\nu} \right) \de_{ab} \nonumber \\
{\cal D}_\om^b\,_c &\equiv& -\, D^\rho \, D_\rho \, \de^b\,_c.
\eeqq

Taking everything together and evaluating the divergent contribution to the one-loop effective action with the use of Eqns.(\ref{138}), (\ref{102}) and (\ref{103}) for $\xi=1$ yields for each independent gauge field and associated ghost
\beq \label{140}
{\it\Delta}_G \! \Ga_{\La, \sl{1}-loop}^{div} \left[A \right]
= - \,\frac{\Om_4}{\ep} \,\frac{1}{6} \,\Om^\La_{\sl 1} 
\, \frac{1}{\La^2} \,\int \,F_{\mu\nu}\,^\al \cdot F^{\mu\nu}\,_\al,
\eeq
where we have discarded the factor $\dim A$ which accounts for the number of independent gauge fields. Note that such a term will reinforce asymptotic freedom. Note in addition that this formula also holds in the Abelian case where the ghost contribution in the presence of $A_\mu\,^\al$ does not reduce to a field-independent determinant.

\subsection{Dirac spinor contribution ${\it\Delta}_D \! \Ga_{\La, \sl{1}-loop}^{div} \left[A \right]$}
The vacuum amplitude of a Dirac field minimally coupled to the gauge field theory of volume-preserving diffeomorphisms is given by
\beq \label{141}
{\cal Z}_D [A] \equiv
\int\Pi_{\!\!\!\!\!\!_{_{_{x,X}}}}\!\!d {\overline\psi} \;
\int\Pi_{\!\!\!\!\!\!_{_{_{x,X}}}}\!\!d \psi
\, \exp\,i\,\Big\{S_D + \ep \mbox{-terms} \Big\}, \eeq
where $\psi$ is a Dirac spinor and 
\beq \label{142}
S_D \equiv – \int \, {\overline\psi} \, \Big({\overline \Dsl} + m \Big) \, \psi
\eeq
is the spinor action coupled to a Yang-Mills field through the covariant derivative $D_\mu [B] = \pa_\mu - i\, t_a B_\mu\,^a$. Here $t_a$ is the generator of the gauge algebra in the fermion space.

Again we have minimally coupled the Dirac field to the gauge field theory of volume-preserving diffeomorphisms replacing ordinary through covariant derivatives $\pa_\mu \ar D_\mu = \pa_\mu + A_\mu\,^\al\cdot \nabla_\al$ yielding
\beq \label{143}
D_\mu [B] \ar {\overline D}_\mu [B] = D_\mu - i\, t_a B_\mu\,^a.
\eeq

Expanding $S_D$ around its stationary points ${\overline\psi}=\psi=B_\mu\,^a = 0$ in the absence of external sources and performing the Grassmann integral gives
\beqq \label{144}
{\cal Z}_{D, \sl{1}-loop} [A]
&=& 
\int\Pi_{\!\!\!\!\!\!_{_{_{x,X}}}}\!\!d\de {\overline\psi} \;
\int\Pi_{\!\!\!\!\!\!_{_{_{x,X}}}}\!\!d\de \psi 
\exp \left\{ - i \, \int \, \de {\overline\psi} \cdot
{\cal D}_\psi \, \de\psi \right\} \nonumber \\
&=& \Det^{1/2}\, {\cal D}^2_\psi, 
\eeqq
where
\beq \label{145}
{\cal D}^2_\psi = - \Dsl^2 =
- D^\rho \, D_\rho \, - \frac{1}{2} \, F^{\mu\nu} \ga_\mu \ga_\nu.
\eeq

Taking everything together and evaluating the divergent contribution to the one-loop effective action with the use of Eqns.(\ref{144}), (\ref{102}) and (\ref{103}) yields for each independent Dirac spinor
\beq \label{146}
{\it\Delta}_D \! \Ga_{\La, \sl{1}-loop}^{div} \left[A \right]
= + \,\frac{\Om_4}{\ep} \,\frac{1}{3} \,\Om^\La_{\sl 1} 
\, \frac{1}{\La^2} \,\int \,F_{\mu\nu}\,^\al \cdot F^{\mu\nu}\,_\al.
\eeq
Note that this will work against asymptotic freedom. Note in addition that a chiral Dirac fields contributes just half of the value above.

\subsection{Scalar doublet contribution ${\it\Delta}_S \! \Ga_{\La, \sl{1}-loop}^{div} \left[A \right]$}
The vacuum amplitude of a complex scalar doublet minimally coupled to the gauge field theory of volume-preserving diffeomorphisms is given by
\beq \label{147}
{\cal Z}_S [A] \equiv
\int\Pi_{\!\!\!\!\!\!_{_{_{x,X}}}}\!\!d \va^\dagger \;
\int\Pi_{\!\!\!\!\!\!_{_{_{x,X}}}}\!\!d \va
\, \exp\,i\,\Big\{S_S + \ep \mbox{-terms} \Big\}, \eeq
where $\va$ is a complex scalar doublet and 
\beq \label{148}
S_S \equiv - \int \, \left( ({\overline D}_\mu \va)^\dagger \cdot ({\overline D}^\mu \va) + V(\va^\dagger\cdot\va) \right)
\eeq
is the doublet coupled to the $SU(2)\times U(1)$ gauge bosons of the electro-weak interaction through the covariant derivative $D_\mu [B] = \pa_\mu - i\, {B\rvec}_\mu\cdot {t\rvec}_\va - i\, B_\mu\, y_\va$.

Again we have minimally coupled the scalar to the gauge field theory of volume-preserving diffeomorphisms replacing ordinary through covariant derivatives $\pa_\mu \ar D_\mu = \pa_\mu + A_\mu\,^\al\cdot \nabla_\al$ yielding
\beq \label{149}
D_\mu [B] \ar {\overline D}_\mu [B] = D_\mu - i\, {B\rvec}_\mu\cdot {t\rvec}_\va - i\, B_\mu\, y_\va.
\eeq

Expanding $S_S$ around one of its stationary points ${B\rvec}_\mu=B_\mu=0$ and  $\va^\dagger\cdot\va=constant$ and performing the Gaussian integral gives
\beqq \label{150}
{\cal Z}_{S, \sl{1}-loop} [A]
&=& 
\int\Pi_{\!\!\!\!\!\!_{_{_{x,X}}}}\!\!d\de \va^\dagger \;
\int\Pi_{\!\!\!\!\!\!_{_{_{x,X}}}}\!\!d\de \va 
\exp \left\{ - i \, \int \, \de \va^\dagger \cdot
{\cal D}_\va \, \de\va \right\} \nonumber \\
&=& \Det^{-1}\, {\cal D}_\va
\eeqq
where
\beq \label{151}
{\cal D}_\va = - D^\rho \, D_\rho \, + \frac{\de V(\va^\dagger\cdot\va)}
{\de\va^\dagger\, \de\va}.
\eeq

Taking everything together and evaluating the divergent contribution to the one-loop effective action with the use of Eqns.(\ref{150}), (\ref{102}) and (\ref{103}) yields for a complex scalar doublet
\beq \label{152}
{\it\Delta}_S \! \Ga_{\La, \sl{1}-loop}^{div} \left[A \right]
= + \,\frac{\Om_4}{\ep} \,\frac{1}{6} \,\Om^\La_{\sl 1} 
\, \frac{1}{\La^2} \,\int \,F_{\mu\nu}\,^\al \cdot F^{\mu\nu}\,_\al,
\eeq
which holds independent of whether the Higgs mechanism is in place or not and will work against asymptotic freedom. Note that a single complex scalar field contributes just half of the value above.

\section{Notations and Conventions}

Generally, ({\bf M}$^{\sl 4}$,\,$\eta$) denotes the four-dimensional Minkowski space with metric $\eta=\mbox{diag}(-1,1,1,1)$, small letters denote space-time coordinates and parameters and capital letters denote coordinates and parameters in inner space.

Specifically, $x^\la,y^\mu,z^\nu,\dots\,$ denote Cartesian space-time coordinates. The small Greek indices $\la,\mu,\nu,\dots$ from the middle of the Greek alphabet run over $\sl{0,1,2,3}$. They are raised and lowered with $\eta$, i.e. $x_\mu=\eta_{\mu\nu}\, x^\nu$ etc. and transform covariantly w.r.t. the Lorentz group $SO(\sl{1,3})$. Partial differentiation w.r.t to $x^\mu$ is denoted by $\pa_\mu \equiv \frac{\pa\,\,\,}{\pa x^\mu}$. Small Latin indices $i,j,k,\dots$ generally run over the three spatial coordinates $\sl{1,2,3}$ \cite{stw1}.

$X^\al, Y^\be, Z^\ga,\dots\,$ denote inner coordinates and $g_{\al\be}$ the flat metric in inner space with signature $-,+,+,+$. The metric transforms as a contravariant tensor of Rank 2 w.r.t. ${\overline{DIFF}}\,{\bf M}^{\sl 4}$. Because Riem$(g) = 0$ we can always globally choose Cartesian coordinates and the Minkowski metric $\eta$ which amounts to a partial gauge fixing to Minkowskian gauges. The small Greek indices $\al,\be,\ga,\dots$ from the beginning of the Greek alphabet run again over $\sl{0,1,2,3}$. They are raised and lowered with $g$, i.e. $x_\al=g_{\al\be}\, x^\be$ etc. and transform as vector indices w.r.t. ${\overline{DIFF}}\,{\bf M}^{\sl 4}$. Partial differentiation w.r.t to $X^\al$ is denoted by $\nabla_\al \equiv \frac{\pa\,\,\,}{\pa X^\al}$. 

The same lower and upper indices are summed unless indicated otherwise.

\end{document}